\documentclass[a4paper]{article}

\usepackage{jheppub} 
                     
\usepackage{graphicx,color}
 \usepackage{bm}
   \usepackage{amsmath}
   \allowdisplaybreaks
    \usepackage{amssymb}    
     \usepackage{pifont}
 \usepackage{stmaryrd}

\usepackage{standalone}
\usepackage{slashed}
\usepackage{multirow}
\usepackage{makecell}
\usepackage{tabu}
\usepackage{hhline}
\usepackage{colortbl}

\newcommand{\nn}{\nonumber}

\newcommand{\Tr}{\mathrm{Tr}}

\renewcommand{\(}{\left(}
\renewcommand{\)}{\right)}
\renewcommand{\[}{\left[}
\renewcommand{\]}{\right]}

\renewcommand{\vec}[1]{\bm{#1}}
\newcommand{\fnot}[1]{\slashed{#1}}

\title{Kinematic power corrections for TMD factorization theorem of semi-inclusive deep-inelastic scattering}

\author{Sara Piloñeta,}

\author{Alexey Vladimirov}

\affiliation{Departamento de F\'isica Te\'orica \& IPARCOS, Universidad Complutense de Madrid, E-28040 Madrid, Spain}

\emailAdd{sarapilo@ucm.es}
\emailAdd{alexeyvl@ucm.es}

\preprint{IPARCOS-UCM-25-052}

\abstract{We evaluate the complete set of kinematic power corrections (KPCs) to the leading power (LP) term of the transverse momentum dependent (TMD) factorization theorem for semi-inclusive deep-inelastic scattering (SIDIS) with a polarized target. This formulation restores the contributions of twist-two TMD distributions to all structure functions, including those that vanish at leading power, such as contributions of longitudinal photons. The resulting expressions are explicitly gauge- and frame-invariant, and inherit all key features of the standard TMD factorization framework, including the coefficient functions and the evolution equations. Numerical estimations show that KPCs contribute only a few percent at $Q\sim10$GeV, but can reach several tens of percents when $Q\sim 2$GeV. Consequently, accounting for kinematic power corrections can be vital for an accurate theoretical description of current SIDIS measurements.}

\begin{document} 
\maketitle
\flushbottom

\section{Introduction}

In recent years, our understanding of the nucleon's internal structure has advanced significantly, partly due to the development of transverse momentum dependent (TMD) factorization theorems \cite{Collins:2011zzd, Echevarria:2011epo, Boussarie:2023izj}. While the predictive power of the TMD factorization approach has been firmly established and its range of successful applications continues to expand, the current leading power (LP) formulation exhibits both theoretical inconsistencies and practical limitations that are becoming progressively apparent (see, for instance, the discussions in refs.~\cite{Angeles-Martinez:2015sea, Gonzalez-Hernandez:2018ipj, Bastami:2018xqd, Ebert:2018gsn, Bacchetta:2019qkv, Hautmann:2020cyp, Ferrera:2023vsw}). It is expected that some of these issues may be resolved, or at least alleviated, through the inclusion of power corrections. Consequently, there has been a growing interest in studying the TMD factorization approach at sub-leading power \cite{Balitsky:2017gis, Ebert:2021jhy, Vladimirov:2021hdn, Balitsky:2021fer, Gamberg:2022lju, Rodini:2022wic, Rodini:2023plb, Vladimirov:2023aot, Balitsky:2024ozy, Arroyo-Castro:2025slx, Jaarsma:2025ksf}.

However, extending the TMD factorization theorem beyond the LP approximation is not a straightforward task, as its structure becomes increasingly complex with each successive power. In contrast to the collinear factorization theorem, where power-suppressed contributions are systematically ordered by powers of the hard scale \cite{Chernyak:1983ej}, the TMD framework involves various kinds of power corrections due to its multi-dimensional nature. 

\begin{figure}
\begin{center}
\includegraphics[width=0.55\textwidth]{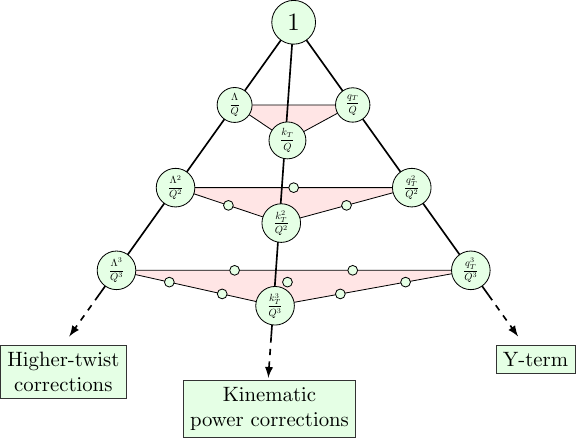}
\end{center}
\caption{\label{fig:power_pyramid} Schematic representation of the ``pyramid" of power corrections in the TMD factorization framework. The levels indicate specific orders in the $1/Q$-expansion, starting with LP at the top. The green bubbles represent power-suppressed terms of different types with three main categories. In the present work, we consider the TMD-with-KPCs approach, which consists in the summation of all KPCs, i.e. all terms along the facing edge.}
\end{figure}

Practically, different types of power corrections correspond to contributions relevant in different kinematic limits of the differential cross-section. Theoretically, they are associated with distinct operator structures. It is convenient to present the hierarchy of power-suppressed terms as a ``pyramid'' of contributions, which naturally emerge from power counting and dimensional analysis within the framework of the TMD operator expansion \cite{Vladimirov:2021hdn}. This idea is visually represented in fig.~\ref{fig:power_pyramid}, where the different tiers correspond to successive orders in the $1/Q$-expansion, i.e., at the apex lies the LP term, followed by the next-to-leading power (NLP) one, and so forth. Continuing with this analogy, the edges host the various types of power corrections: higher-twist contributions (denoted as $\Lambda/Q$), $q_T/Q$ terms (often associated with the Y-term \cite{Collins:2011zzd}) and kinematic power corrections (KPCs) (marked as $k_T/Q$). This structure becomes evident already at NLP, whose expression can be written as a sum of three independent contributions with specific properties \cite{Arroyo-Castro:2025slx}. For higher power terms, one also expects mixed-type contributions, indicated by the small dots in fig.~\ref{fig:power_pyramid}. Each class of power corrections exhibits distinct features, and the interplay among the relevant physical scales makes certain kinds more significant than others, depending on the kinematic regime under consideration. For a more detailed discussion, we refer to ref.~\cite{Vladimirov:2023aot}.

Among the various types of power corrections, KPCs play a special role, and posses unique properties. They encompass all power-suppressed terms involving twist-two TMD distributions at small $q_T$, restore both gauge- and frame-invariance of the hadronic tensor, and structurally resemble the LP term, sharing its non-perturbative content and coefficient functions. For all these reasons, incorporating KPCs into the TMD factorization theorem seems natural and advantageous. Such an extension has already been carried out in ref.~\cite{Vladimirov:2023aot}, where the KPCs following the LP term were summed and derived at all powers. This improved formalism is referred to as the TMD-with-KPCs factorization theorem, and can be seen as an extension of the ordinary LP TMD factorization into an explicitly Lorentz and gauge invariant form, while preserving its original properties. The TMD-with-KPC approach disregards the strict $Q$-counting, replacing it by the TMD-twist-counting and allowing to determine the leading-twist contribution to all observables (in this sense, it is similar to the computation of Wandzura-Wilczek terms \cite{Bastami:2018xqd} at all powers). Its domain of applicability coincides with that of the LP TMD factorization theorem (that is, $q_T \ll Q$ and $\Lambda \ll Q$). The practical feasibility of this framework has already been demonstrated in \cite{Piloneta:2024aac}, where all angular distributions that appear in the Drell–Yan (DY) lepton pair production, even the sub-leading ones, were successfully described.

In this work, we consider the structure functions of semi-inclusive deep inelastic scattering (SIDIS) and derive their expressions within the TMD-with-KPCs formalism. The LP and NLP expressions for these structure functions are already known \cite{Bacchetta:2006tn, Ebert:2021jhy, Rodini:2023plb}. However, at this level they explicitly violate frame- and charge-conservation. While such violations may be negligible at asymptotically large $Q$, they are not at the scales relevant for modern SIDIS measurements with $Q\sim 1-5$GeV. By resumming KPCs, we restore these critical symmetries, and, in addition, gain access to the leading-twist approximation of sub-leading structure functions, such as those induced by longitudinally polarized photons. These improvements are expected to enhance the description of SIDIS and help resolve discrepancies observed in phenomenological studies \cite{Bacchetta:2017gcc, Scimemi:2019cmh, Bacchetta:2022awv,Bacchetta:2024qre, Moos:2025sal}. This issue becomes particularly critical in view of the preparation of precise SIDIS data for future colliders  \cite{AbdulKhalek:2021gbh, AbdulKhalek:2022hcn, Accardi:2023chb}.

The article is organized as follows: in section~\ref{sec:theory} we present the theoretical framework and summarize the main aspects of SIDIS kinematics (sec.~\ref{subsec:SIDIS_kinematics}) that are relevant for computing the two key theoretical ingredients: the lepton tensor, discussed in sec.~\ref{subsec:lepton_tensor}, and the hadron tensor. The later is derived in sec.~\ref{subsec:hadron_tensor} within the TMD-with-KPCs factorization theorem, where we also outline the main features of this formalism. The resulting expressions for the structure functions are presented in sec.~\ref{subsec:StructFunctions}. Finally, in section~\ref{sec:impact_studies}, we explore the phenomenological implications of including KPCs using the \texttt{artemide} code \cite{Scimemi:2019cmh} together with the ART25 set of TMD distributions \cite{Moos:2025sal}. Additional relations useful for the structure functions computation are collected in appendices~\ref{app:frames_dictionary} and \ref{app:largeQ}, while appendix~\ref{app:convolution_integral} provides the technical details of the momentum-space convolution and its implementation in \texttt{artemide}.

\section{Semi-Inclusive Deep Inelastic scattering}
\label{sec:theory}

In this work, we focus on the SIDIS process, which is defined by the reaction
\begin{eqnarray}\label{eq:sidis_reaction}
    \ell(l) + H(P) \rightarrow \ell(l') + h(p_h) + X,
\end{eqnarray}
where $\ell$ is the incoming lepton, $H$ and $h$ represent the target and produced hadrons, respectively, and $X$ denotes the collection of undetected final states. The four-momentum of each particle is indicated in parentheses. In the following, we treat the lepton as massless, i.e., $l^2 = l'^2 = m_l^2 = 0$, while the masses of the hadrons are given by $P^2 = M^2$ and $p_h^2 = m_h^2$.

A total of 18 structure functions are employed to parametrize polarized SIDIS, each proportional to sine and cosine modulations of various azimuthal angles \cite{Bacchetta:2006tn}. Although deriving these functions from a given model is, in principle, a straightforward task, the computation can become intricate due to the numerous scalar products, whose explicit formulas are often lengthy. The main algebraic complication arises from the fact that the angles are defined with respect to the lepton in the laboratory frame, while the factorization is performed in a system where the momenta of the hadrons are characterized in a more natural way. Reconciling these two reference frames to resolve the resulting mismatch leads to cumbersome intermediate expressions. For a review, see, e.g., refs.~\cite{Diehl:2005pc, Bastami:2018xqd, Boer:2011xd, Boglione:2019nwk,  Ebert:2021jhy, Rodini:2023plb}, where different approaches are discussed. In this article, we outline an alternative formulation, in which each structure function is presented as a product of two tensors explicitly constructed in the hadron frame. This representation is conceptually close to the one used in \cite{Diehl:2005pc, Ebert:2021jhy}, but employs a basis of physical vectors instead of the more abstract helicity basis. It should be noted that an analogous decomposition for the Drell-Yan reaction was previously described in ref.~\cite{Piloneta:2024aac}.

We assume that the interaction between the lepton and the nucleon target takes place through the exchange of a single virtual photon. Under this assumption, the differential cross-section of the process in (\ref{eq:sidis_reaction}) reads
\begin{eqnarray}\label{eq:sidis_initial_dsigma}
    d\sigma = \frac{2}{s-M^2}\frac{\alpha_{\rm{em}}^2}{(q^2)^2}L_{\mu\nu}W^{\mu\nu}\frac{d^3l'}{2E'}\frac{d^3p_h}{2E_h},
\end{eqnarray}
where $s$ is the center-of-mass energy, $\alpha_{\rm{em}} = e^2/4\pi$ is the electromagnetic coupling constant, and $q = l - l'$ indicates the momentum of the mediating photon. Accordingly, the $q^2$ factors arise from the photon propagator $\Delta^{\mu\nu} = g^{\mu\nu}/(q^2 + i0)$. The last terms in eq.~(\ref{eq:sidis_initial_dsigma}) correspond to the phase-space elements of the detected lepton and the produced hadron, where $E'$ and $E_h$ denote their respective energies. 

To compute the cross-section, it is necessary to evaluate both the lepton ($L^{\mu\nu}$) and hadron ($W^{\mu\nu}$) tensors. Before proceeding with their explicit computation, however, it is useful to review the main kinematic features of the SIDIS process. In this section, we summarize its kinematics and algebraic structure, and derive convenient pocket formulas for its structure functions. Note that the calculations presented here are fully general, and therefore we keep the hadrons masses non-zero, although they will be neglected in the following sections devoted specifically to TMD factorization. 

\subsection{Brief overview of SIDIS kinematics}
\label{subsec:SIDIS_kinematics}

The kinematic variables relevant for the description of SIDIS are defined as
\begin{eqnarray}\label{eq:sidis_variables}
    Q^2 = -q^2, \quad x = \frac{Q^2}{2(Pq)}, \quad y = \frac{(Pq)}{(Pl)}, \quad z = \frac{(P p_h)}{(P q)}.
\end{eqnarray}
Here, $Q^2$ denotes the squared momentum transferred to the incoming lepton, equal to the virtuality of the mediating photon. The Bjorken variable $x$ specifies the fraction of the nucleon's momentum carried by the struck parton, while the inelasticity $y$ determines how much energy the lepton deposits into the target in the laboratory frame. The variable $z$ represents the portion of the virtual photon's momentum taken by the detected hadron. It is useful to recall that these invariants are related to each other through the Mandelstam variable $s=(P+l)^2$ as
\begin{eqnarray}
    xy(s-M^2)=Q^2.
\end{eqnarray} 
We also introduce the following set of power-suppressed variables,
\begin{eqnarray}
    \gamma = \frac{2Mx}{Q}, \quad \gamma_h = \frac{m_h}{zQ},
\end{eqnarray}
which typically appear in the treatment of target- and produced-mass corrections.

To obtain a complete description of hadron structure, one must account for the transverse momentum exchanged between the colliding partons. This can be achieved by parameterizing the underlying kinematic quantities in different vector bases. In practice, two reference frames are generally employed. The first defines transverse components with respect to the vectors $P^\mu$ and $q^\mu$, and represents the conventional choice used in experimental analyses, where all observables are formulated. It is denoted by the $\perp$-symbol. The second frame is constructed relative to $P^\mu$ and $p_h^\mu$, providing a natural setting for theoretical computations, as the partons' transverse momenta are expressed in this basis. It is marked by the $T$-symbol. Below, we review the significant vector definitions in both systems and discuss their relations.

\subsubsection{The transverse $\perp$-plane}\label{subsubsec:perp_plane}

We begin with the frame in which the experimentally accessible quantities are defined, commonly referred to as the ``laboratory frame''. Typically, what is measured is the transverse component of the detected hadron momentum $p_h^\mu$ with respect to the plane spanned by the vectors $P^\mu$ and $q^\mu$. Consequently, the components orthogonal to this plane, traditionally denoted by the $\perp$-symbol, can be obtained using the projector
\begin{eqnarray}
    g_\perp^{\mu\nu} = g^{\mu\nu} - \frac{1}{Q^2(1+\gamma^2)}\left[4x^2P^\mu P^\nu - \gamma^2q^\mu q^\nu + 2x(P^\mu q^\nu+q^\mu P^\nu)\right],
\end{eqnarray}
normalized such that $g_\perp^{\mu\nu}g_{\perp,\mu\nu} = 2$. Additionally, a second projector that orthogonalizes a transverse vector can be constructed with the aid of the transverse Levi-Civita tensor\footnote{
The Minkowski Levi-Civita tensor is defined as $\epsilon^{0123}=+1$. All sign conventions used in this work coincide with \cite{Bacchetta:2006tn}.
}
\begin{equation}
\epsilon_\perp^{\mu\nu} = \frac{2x}{Q^2\sqrt{1+\gamma^2}}\epsilon^{\mu\nu\rho\sigma}P_\rho q_\sigma,
\end{equation}
which satisfies $\epsilon_\perp^{\mu\rho}\epsilon_{\perp\,\rho}^{\nu}=g_\perp^{\mu\nu}$. 

\begin{figure}
\begin{center}
\includegraphics[width=0.55\textwidth]{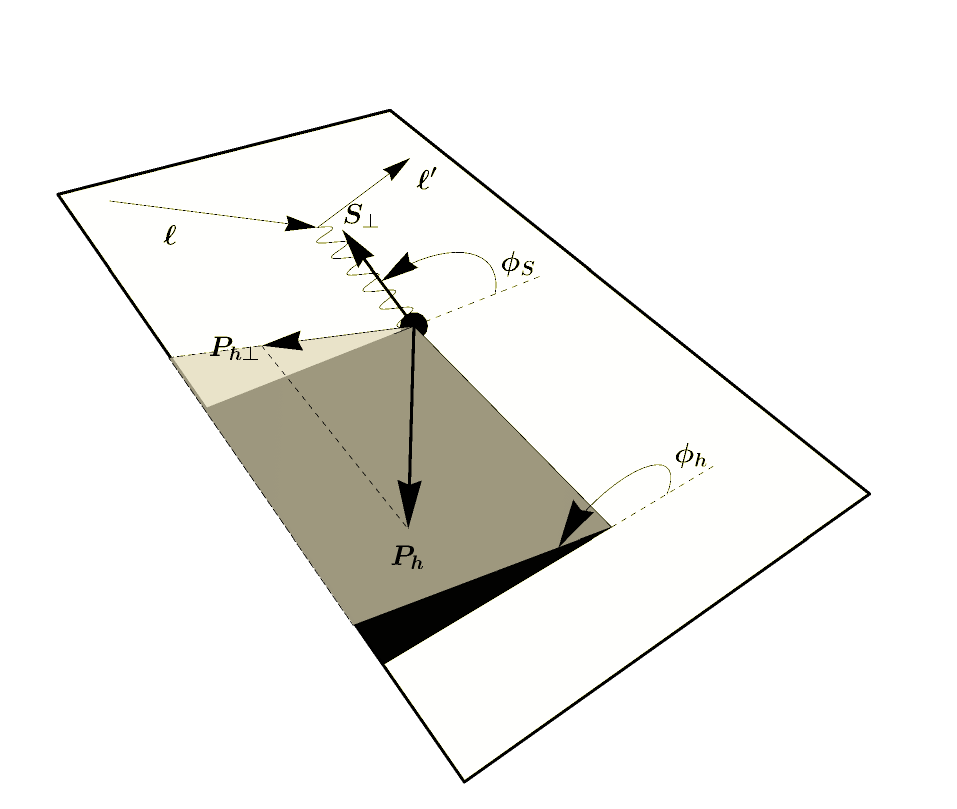}
\end{center}
\caption{\label{fig:sidis_planes} Description of the azimuthal angles according to the Trento conventions \cite{Bacchetta:2004jz}.}
\end{figure}

These two projectors allow the construction of an orthogonal basis in the transverse $\perp$-plane. Accordingly, the transverse component of a generic four-vector $v^\mu$ is given by $v^\mu_\perp=g_\perp^{\mu\nu}v_\nu$, while the transverse (axial) vector orthogonal to $v_\perp^\mu$ is denoted with a tilde and determined by $\tilde v_\perp^\mu=\epsilon^{\mu\nu}_\perp v_\nu$. Vectors belonging to the $\perp$-plane are purely spatial, and thus have a negative Minkowski scalar product. In what follows, we adopt boldface to denote the magnitude of this product, i.e. $\vec{v}_\perp^2=-g_\perp^{\mu\nu}v_\mu v_\nu=-v_\perp^2>0$. In summary, the notation reads
\begin{eqnarray}
v_\perp^\mu=g^{\mu\nu}_\perp v_\nu,\qquad
\tilde v_\perp^\mu=\epsilon^{\mu\nu}_\perp v_\nu, \qquad
(v_\perp \cdot \tilde v_\perp)=0,\qquad \vec v_\perp^2=-v_\perp^2=-\tilde v_\perp^2>0.
\end{eqnarray}
On the other hand, the azimuthal angles are defined with respect to the lepton momentum, following the Trento conventions \cite{Bacchetta:2004jz} (see fig.~\ref{fig:sidis_planes}), and can be computed using the tensors $g_\perp^{\mu\nu}$ and $\epsilon_\perp^{\mu\nu}$ introduced above. In particular, the azimuthal angle for the produced hadron is
\begin{equation}\label{eq:hadron_azimuthal_angle}
\cos{\phi_h} = -\frac{l_\mu p_{h\nu}g_\perp^{\mu\nu}}{\sqrt{\bm{l_\perp}^2\bm{p_\perp}^2}}, \qquad
\sin{\phi_h} = -\frac{l_\mu p_{h\nu}\epsilon_\perp^{\mu\nu}}{\sqrt{\bm{l_\perp}^2\bm{p_\perp}^2}}.
\end{equation}
Similarly, the azimuthal angle associated with the spin vector $S^\mu$ is defined analogously to $\phi_h$ as
\begin{eqnarray}
    \cos{\phi_S}=-\frac{l_\mu S_\nu g^{\mu\nu}_\perp}{\sqrt{\bm{l}^2_\perp\bm{S}^2_\perp}}, \qquad \sin{\phi_S}=-\frac{l_\mu S_\nu \epsilon^{\mu\nu}_\perp}{\sqrt{\bm{l}^2_\perp\bm{S}^2_\perp}}.
\end{eqnarray}
The relative azimuthal angle between the hadron and spin directions, $\phi_h - \phi_S$, is therefore given by
\begin{eqnarray}
    \cos{(\phi_h-\phi_S)}=-\frac{S_\mu p_{h,\nu}g^{\mu\nu}_\perp}{\sqrt{\bm{p}^2_\perp\bm{S}^2_\perp}}, \qquad \sin{(\phi_h-\phi_S)}=-\frac{S_\mu p_{h,\nu}\epsilon^{\mu\nu}_\perp}{\sqrt{\bm{p}^2_\perp\bm{S}^2_\perp}}.
\end{eqnarray}
In this way, the $\perp$-coordinate system is entirely specified by the set of vectors $(P^\mu,q^\mu,p^\mu_\perp,\tilde{p}^\mu_\perp)$. Any other vector can be expressed in this system in terms of the variables $(x,z,Q,\vec p_\perp^2)$ and the relevant azimuthal angles. Specifically, the decomposition of the spin vector takes the form
\begin{eqnarray}\label{def:Smu}
S^\mu = S_\parallel \frac{2xP^\mu-\gamma^2q^\mu}{Q\gamma\sqrt{1+\gamma^2}}+|\bm{S}_\perp|\frac{\cos{(\phi_h-\phi_S)}p^\mu_\perp+\sin{(\phi_h- \phi_S)}\tilde{p}^\mu_\perp}{\sqrt{\bm{p}^2_\perp}},
\end{eqnarray}
where the longitudinal component is determined by
\begin{eqnarray}
S_\|=\frac{(Sq)}{Q}\frac{\gamma}{\sqrt{1+\gamma^2}},
\end{eqnarray}
and $\vec S_\perp$ denotes the perpendicular one.

\subsubsection{The transverse $T$-plane}\label{subsubsec:T_plane}

The TMD factorization for SIDIS is most conveniently formulated in the so-called Breit frame (occasionally referred to as ``the factorization frame'' throughout this work), where the momenta of the involved hadrons are approximately light-like and aligned in opposite directions. This configuration naturally suggests the parametrization
\begin{eqnarray}\label{eq:hadrons_momenta}
    P^\mu = P^+\bar{n}^\mu + \frac{M^2}{2P^+}n^\mu, \qquad p_h^\mu = p_h^-n^\mu + \frac{m_h^2}{2p_h^-}\bar{n}^\mu,
\end{eqnarray}
where $n^\mu$ and $\bar{n}^\mu$ are two independent light-like vectors normalized such that $(n\bar{n}) = 1$. 

We adopt the standard notation for the light-cone decomposition of a generic four-vector,
\begin{eqnarray}
    v^\mu = v^+\bar{n}^\mu + v^-n^\mu + v_T^\mu,
\end{eqnarray}
with $v^+ = (nv)$, $v^- = (\bar{n}v)$ and $v_T^\mu$ denoting the transverse component orthogonal to the $(n,\bar{n})$-plane, satisfying $v_T^2 < 0$. Thus, the label $T$ refers to transverse components defined in the hadron frame, which should not be confused with the transverse $\perp$-components of the hadron-photon frame.

The symmetric and anti-symmetric projectors onto the transverse $T$-plane are
\begin{eqnarray}\label{eq:gTmunu_nnbar}
g_T^{\mu\nu} &=& g^{\mu\nu} - n^\mu\bar{n}^\nu - \bar{n}^\mu n^\nu,
\\\label{eq:eps_T}
\epsilon_T^{\mu\nu} &=& \epsilon^{\mu\nu-+}=\epsilon^{\mu\nu\alpha\beta}\bar n_\alpha n_\beta,
\end{eqnarray}
normalized such that $g_T^{\mu\nu}g_{T,\mu\nu}=2$ and $\epsilon_T^{\mu\rho}\epsilon_{T\,\rho}^\nu=g_T^{\mu\nu}$. Analogously to the $\perp$-system, we define $v_T^\mu$ as the vector orthogonal to $\tilde v_T^\mu$. The notation for the $T$-system is then summarized as
\begin{eqnarray}
v_T^\mu=g^{\mu\nu}_T v_\nu,\qquad
\tilde v_T^\mu=\epsilon^{\mu\nu}_T v_\nu, \qquad
(v_T \cdot \tilde v_T)=0,\qquad \vec v_T^2=-v_T^2=-\tilde v_T^2>0,
\end{eqnarray}
where boldface denotes the magnitude of the scalar product for purely transverse vectors.

It should be noted that, by definition of (\ref{eq:hadrons_momenta}), the $(n,\bar{n})$-plane coincides with the one spanned by the hadron momenta $P^\mu$ and $p_h^\mu$. Therefore, the light-cone vectors can be directly expressed in terms of the kinematic variables introduced earlier,
\begin{eqnarray}\label{def:n}
n^\mu &=& \frac{2xP^+}{zQ^2\sqrt{1-\gamma^2\gamma_h^2}}\left(p_h^\mu-P^\mu\frac{2xz(1-\sqrt{1-\gamma^2\gamma_h^2})}{\gamma^2}\right), 
\\\label{def:nBar}
\bar{n}^\mu &=& \frac{1}{2P^+\sqrt{1-\gamma^2\gamma_h^2}}\left(P^\mu(1+\sqrt{1-\gamma^2\gamma_h^2})-p_h^\mu\frac{\gamma^2}{2xz}\right).
\end{eqnarray}
Here, we have used the relation
\begin{eqnarray}
P^+p_h^-=Q^2 \frac{z}{4x}\(1+\sqrt{1-\gamma^2 \gamma_h^2}\).
\end{eqnarray}
The component $P^+$ remains unspecified; however, it cancels out in all Lorentz invariant quantities.

Using the equations (\ref{def:n}, \ref{def:nBar}) we can rewrite the projectors (\ref{eq:gTmunu_nnbar}, \ref{eq:eps_T}) in physical terms
\begin{eqnarray}
g_T^{\mu\nu} &=& g^{\mu\nu} - \frac{1}{Q^2(1-\gamma^2\gamma_h^2)}\left[-4x^2\gamma_h^2 P^\mu P^\nu - \frac{\gamma^2}{z^2}p_h^\mu p_h^\nu + \frac{2x}{z}(P^\mu p_h^\nu+p_h^\mu P^\nu)\right],
\\
\epsilon_T^{\mu\nu} &=& \frac{2x}{zQ^2\sqrt{1-\gamma^2\gamma_h^2}}\epsilon^{\mu\nu\rho\sigma}P_\sigma p_{h\rho}.
\end{eqnarray}
These two tensors allow us to express any vector in terms of $(P^\mu,p_h^\mu, q_T^\mu, \tilde q_T^\mu)$ or, alternatively, in terms of $(n^\mu, \bar n^\mu, q_T^\mu, \tilde q_T^\mu)$. In particular, the target spin vector can be written as follows
\begin{eqnarray}
S^\mu = \lambda \frac{P^+\bar{n}^\mu-P^-n^\mu}{M}+S^\mu_T = \lambda \frac{2x z P^\mu-\gamma^2 p_h^\mu}{z\gamma Q\sqrt{1-\gamma^2 \gamma_h^2}}+S_T^\mu,
\end{eqnarray}
where $\lambda$ is the longitudinal spin component and $S^\mu_T$ denotes the transverse one. 

Let us remark that, in the absence of masses, the vectors $P^\mu$ and $p_h^\mu$ tend to $\bar n^\mu$ and $n^\mu$, respectively. Therefore, one set of notation directly matches another. Since target mass corrections to TMD factorization are currently unknown, the vectors $n$ and $\bar n$ are natural for theoretical computations. Thus, along this work, we often refer to the systems $(n,\bar n)$ and $(P,p_h)$ synonymously.

\subsubsection{Relation between systems}

In order to bridge the two vector systems introduced above, it is helpful to establish a set of relations that facilitate transitions between them without undue complications. This is accomplished by comparing Lorentz-invariant quantities, such as scalar products, expressed in both frames. Adopting this strategy, we have constructed a ``dictionary'' of helpful relations between variables in the $T$- and $\perp$-systems, including also some scalar products that have proven useful in our computations. For instance, calculating $\vec q_T^2$ and $\vec p_\perp^2$ using the corresponding projectors yields the relation
\begin{eqnarray}\label{rel:qT-to-pPerp}
    \vec q_T^2 = \frac{\vec p_\perp^2}{z^2}\frac{1+\gamma^2}{1-\gamma^2\gamma_h^2},
\end{eqnarray}
which is particularly important since it connects the transverse momentum appearing in the TMD formalism, $q_T$, with the one actually measured in experiments, $p_\perp$.  The remaining entries of this dictionary are provided in appendix \ref{app:frames_dictionary}. 

\subsection{Cross-section and structure functions definition}

The differential cross-section (\ref{eq:sidis_initial_dsigma}) can be fully expressed in terms of the kinematic variables introduced so far. To achieve this, we need the explicit expressions for the phase-space elements of the outgoing lepton and the detected hadron. These can be obtained by decomposing their four-momenta and projecting onto the relevant angular directions. The full derivation is available in the literature and will not be reproduced here. The resulting expressions are
\begin{eqnarray}\label{eq:phase_space_elements}
    \frac{d^3l'}{2E'} = \frac{y}{4x}dxdQ^2d\psi, \quad     \frac{d^3p_h}{2E_h} = \frac{1}{4z\sqrt{1-\gamma^2\gamma_h^2-\gamma^2\frac{\vec p_\perp^2}{z^2Q^2}}}\,dzd\vec p_\perp^2d\phi_h,
\end{eqnarray}
where $\phi_h$ is the azimuthal angle defined in (\ref{eq:hadron_azimuthal_angle}), and $\psi$ is the azimuthal angle of the outgoing lepton around the lepton beam axis (see fig.~\ref{fig:sidis_planes}), measured with respect to an arbitrary fixed direction. The angle $\psi$ is related to the $\phi_S$ as \cite{Diehl:2005pc}
\begin{eqnarray}
\cos\psi=\frac{\cos\phi_S}{\sqrt{1-\sin^2\theta_\gamma \sin^2\phi_S}},
\end{eqnarray}
where $\sin\theta_\gamma\sim \gamma$ (see definition in eqn.~(\ref{eq:theta-gamma})). Therefore, in the $Q\gg M$ limit, one finds $\psi\approx \phi_S$.

By inserting eqn.~(\ref{eq:phase_space_elements}) into eqn.~(\ref{eq:sidis_initial_dsigma}) we obtain the following expression for the SIDIS differential cross-section
\begin{eqnarray}\label{eq:sidis_differential_cross_section}
    \frac{d\sigma}{dxdyd\psi dzd\phi_hd\bm{p_\perp}^2} = \frac{\alpha_{em}^2}{Q^4}\frac{y}{8z}\frac{L_{\mu\nu}W^{\mu\nu}}{\sqrt{1-\gamma^2\gamma_h^2-\gamma^2\frac{\bm{p_\perp}^2}{z^2Q^2}}},
\end{eqnarray}
which reproduces the well-known result \cite{Diehl:2005pc, Bacchetta:2006tn}.

The SIDIS cross-section is parameterized in terms of 18 independent structure functions \cite{Bacchetta:2006tn}
\begin{eqnarray}
\nn
\frac{d\sigma}{dxdyd\psi dzd\phi_h d\vec p_\perp^2}&=&\frac{\alpha^2_{\text{em}}}{xyQ^2}\frac{y^2}{2(1-\varepsilon)}\(1+\frac{\gamma^2}{2x}\)\Bigg\{F_{UU,T}+\varepsilon F_{UU,L}+\sqrt{2\varepsilon(1+\varepsilon)}\cos\phi_h F_{UU}^{\cos \phi_h}
\\\nn &&
+\varepsilon \cos(2\phi_h)F_{UU}^{\cos2\phi_h}+\lambda_e\sqrt{2\varepsilon(1-\varepsilon)}\sin\phi_hF_{LU}^{\sin\phi_h}
\\\nn && 
+S_\|\(\sqrt{2\varepsilon(1+\varepsilon)}\sin\phi_h F_{UL}^{\sin\phi_h}+\varepsilon \sin(2\phi_h)F_{UL}^{\sin2\phi_h}\)
\\\nn && 
+S_\|\lambda_e\(\sqrt{1-\varepsilon^2}F_{LL}+\sqrt{2\varepsilon(1-\varepsilon)} \cos \phi_h F_{LL}^{\cos\phi_h}\)
\\
\label{def:structure-functions}
&& 
+|\vec S_\perp|\Big(\sin(\phi_h-\phi_S)(F_{UT,T}^{\sin(\phi_h-\phi_S)}+\varepsilon F_{UT,L}^{\sin(\phi_h-\phi_S)})
\\\nn &&
\qquad +\varepsilon \sin(\phi_h+\phi_S)F_{UT}^{\sin(\phi_h+\phi_S)}
+\varepsilon \sin(3\phi_h-\phi_S)F_{UT}^{\sin(3\phi_h-\phi_S)}
\\\nn &&
\qquad +\sqrt{2\varepsilon(1+\varepsilon)}\sin \phi_S F_{UT}^{\sin \phi_S}
+\sqrt{2\varepsilon(1+\varepsilon)}\sin(2\phi_h-\phi_S) F_{UT}^{\sin(2\phi_h-\phi_S)}
\Big)
\\\nn && 
+|\vec S_\perp|\lambda_e\Big(\sqrt{1-\varepsilon^2}\cos(\phi_h-\phi_S)F_{LT}^{\cos(\phi_h-\phi_S)}
+\sqrt{2\varepsilon(1-\varepsilon)}\cos \phi_S F_{LT}^{\cos \phi_S}
\\\nn && \qquad
+\sqrt{2\varepsilon(1-\varepsilon)}\cos(2\phi_h-\phi_S) F_{LT}^{\cos(2\phi_h-\phi_S)}
\Big)\Bigg\},
\end{eqnarray}
where $\lambda_e$ is the helicity of the lepton and $\varepsilon$ is the ratio of 
longitudinal and transverse photon flux,
\begin{eqnarray}\label{def:varepsilon}
\varepsilon = \frac{1-y-\frac{1}{4}\gamma^2y^2}{1-y+\frac{1}{2}y^2+\frac{1}{4}\gamma^2y^2}.
\end{eqnarray}
The super-scripts of the structure functions introduced in (\ref{def:structure-functions}) encode their azimuthal dependence, while the sub-scripts specify the polarization of the lepton and the target hadron, respectively. In particular, the first one, $U$ ($L$), refers to an unpolarized (longitudinally polarized) lepton beam, whereas the second, $U$ ($L$ or $T$), corresponds to an unpolarized (longitudinally or transversely polarized) target hadron, with polarization defined relative to the virtual photon direction. In addition, a third subscript appears in four of the structure functions and indicates the polarization of the exchanged photon. 

Additionally, it should be noted that the factor $(1+\gamma^2/2x)$, included in the standard definition of structure functions \cite{Bacchetta:2006tn}, does not arise from any invariant decomposition. Therefore, following the standard convention, we also introduce it in eqn.~(\ref{def:structure-functions}); however, it leads to a compensating factor within the theoretical expressions for the structure functions derived below.

\subsection{Lepton tensor}
\label{subsec:lepton_tensor}

The lepton tensor reads
\begin{eqnarray}\label{eq:lepton_tensor}
L_{\mu\nu} = 2(l_\mu l'_\nu + l'_\mu l_\nu - (l l')g_{\mu\nu}) + 2i\lambda_e \epsilon_{\mu\nu\rho\sigma}l^\rho l'^\sigma.
\end{eqnarray}
To make the azimuthal angle dependence explicit, we rewrite it in the $\perp$-system, where the lepton momentum takes the form
\begin{eqnarray}\label{eq:lepton_momentum_decomposition}
l^\mu = \frac{x}{y}\frac{2-y}{1+\gamma^2}P^\mu + \frac{2 + y\gamma^2}{2y(1+\gamma^2)}q^\mu + \frac{Q}{y\gamma}\cos{\phi_h}\sin{\theta_\gamma}\frac{p_\perp^\mu}{\sqrt{\bm{p_\perp}^2}} + \frac{Q}{y\gamma}\sin{\phi_h}\sin{\theta_\gamma}\frac{\tilde{p}_\perp^\mu}{\sqrt{\bm{p_\perp}^2}}.
\end{eqnarray}
Here, $\theta_\gamma$ denotes the angle between $l$ and $q$ in the laboratory gamma-hadron frame \cite{Kotzinian:1994dv}, defined as
\begin{eqnarray}\label{eq:theta-gamma}
    \cos{\theta_\gamma} = \frac{1+y\frac{\gamma^2}{2}}{\sqrt{1+\gamma^2}} , \quad \sin{\theta_\gamma} = \frac{\gamma}{\sqrt{1+\gamma^2}}\sqrt{1-y-y^2\frac{\gamma^2}{4}}.
\end{eqnarray} 
Substituting eqn.~(\ref{eq:lepton_momentum_decomposition}) into eqn.~(\ref{eq:lepton_tensor}) yields the following decomposition 
\begin{eqnarray}\label{eq:lepton_tensor_decomposition}
L^{\mu\nu} &=& \frac{Q^2}{1-\varepsilon}\Big[-\mathcal{S}_0^{\mu\nu} + (1+2\varepsilon)\mathcal{S}_1^{\mu\nu} + \lambda_e\sqrt{1-\varepsilon^2}\mathcal{A}_1^{\mu\nu} \\ \nn &&
+ \sqrt{2\varepsilon(1+\varepsilon)}\cos{\phi_h}\mathcal{S}_3^{\mu\nu} + \lambda_3\sqrt{2\varepsilon(1-\varepsilon)}\cos{\phi_h}\mathcal{A}_2^{\mu\nu} + \sqrt{2\varepsilon(1+\varepsilon)}\sin{\phi_h}\mathcal{S}_5^{\mu\nu} \\ \nn &&
\lambda_e\sqrt{2\varepsilon(1-\varepsilon)}\sin{\phi_h}\mathcal{A}_3^{\mu\nu} + \varepsilon\cos{2\phi_h}(\mathcal{S}_0^{\mu\nu}-\mathcal{S}_1^{\mu\nu}-2\mathcal{S}_2^{\mu\nu}) - \varepsilon\sin{2\phi_h}\mathcal{S}_4^{\mu\nu}\Big], 
\end{eqnarray}
where $\mathcal{S}$ and $\mathcal{A}$ denote symmetric and anti-symmetric projectors, respectively. Their explicit expressions are
\begin{align}\label{def:S0}
    \mathcal{S}_0^{\mu\nu} &= g^{\mu\nu} + \frac{q^\mu q^\nu}{Q^2} , \\
    \mathcal{S}_1^{\mu\nu} &= \frac{(2xP^\mu + q^\mu)(2xP^\nu + q^\nu)}{(1+\gamma^2)Q^2} , \\
    \mathcal{S}_2^{\mu\nu} &= \frac{p_\perp^\mu p_\perp^\nu}{-\bm{p_\perp}^2} , \\
    \mathcal{S}_3^{\mu\nu} &= \frac{(2xP^\mu + q^\mu)p_\perp^\nu + p_\perp^\mu(2xP^\nu + q^\nu)}{(1+\gamma^2)Q\sqrt{\bm{p_\perp}^2} }, \\
    \mathcal{S}_4^{\mu\nu} &= \frac{p_\perp^\mu\tilde{p}_\perp^\nu + \tilde{p}_\perp^\mu p_\perp^\nu}{-\bm{p_\perp}^2} , \\
    \mathcal{S}_5^{\mu\nu} &= \frac{(2xP^\mu + q^\mu)\tilde{p}_\perp^\nu + \tilde{p}_\perp^\mu(2xP^\nu + q^\nu)}{(1+\gamma^2)Q\sqrt{\bm{p_\perp}^2}} , \\
    \mathcal{A}_1^{\mu\nu} &=  -i\epsilon_\perp^{\mu\nu }, \\
    \mathcal{A}_2^{\mu\nu} &= -i\frac{\epsilon^{\mu\nu\rho\sigma}p_{\perp\sigma}q_\rho}{Q\sqrt{\bm{p}_\perp^2}}  , \\ \label{def:A3}
    \mathcal{A}_3^{\mu\nu} &= i \frac{(2xP^\mu + q^\mu)p_\perp^\nu - p_\perp^\mu(2xP^\nu + q^\nu)}{(1+\gamma^2)Q\sqrt{\bm{p_\perp}^2}} .
\end{align}
These tensors are dimensionless and transverse to $q^\mu$, i.e. $q_\mu \mathcal{S}_i^{\mu\nu}=q_\mu \mathcal{A}_i^{\mu\nu}=0$. It is worth noting that $\mathcal{S}_4$, $\mathcal{S}_5$, $\mathcal{A}_1$, and $\mathcal{A}_2$ are \textit{P}-odd, while all others are \textit{P}-even. Moreover, these projectors remain regular in the massless limit, which is obtained by setting $\gamma\to0$.

\subsection{Structure functions pocket formulas}

The hadron tensor is at most linear in the spin-vector (\ref{def:Smu}). Therefore, it can be written as
\begin{eqnarray}\label{eq:hadron_tensor_spin_decomposition}
W^{\mu\nu} = W_U^{\mu\nu} + S_\parallel W_L^{\mu\nu} + |\vec S_\perp|\cos(\phi_h-\phi_S)W_S^{\mu\nu} + |\vec S_\perp|\sin(\phi_h-\phi_S)W_A^{\mu\nu} .
\end{eqnarray}
Here, the tensors $W_U$ and $W_A$ are \textit{P}-even, while $W_L$ and $W_S$ are \textit{P}-odd.

Contracting the hadronic tensor (\ref{eq:hadron_tensor_spin_decomposition}) with the lepton tensor (\ref{eq:lepton_tensor_decomposition}), we obtain an expression for the cross-section with the azimuthal angle structure explicitly displayed. To facilitate this procedure, it is convenient to take into account the parity properties of the various tensor components. By comparing the result with the parameterization in (\ref{def:structure-functions}), each structure function can then be identified with a convolution of particular tensors. The resultant formulas are the following
\begin{eqnarray}\label{def:FUUT}
&& F_{UU,T} = \frac{x}{4z}F_0 (\mathcal{S}_1^{\mu\nu}-\mathcal{S}_0^{\mu\nu})W_{U,\mu\nu} , 
\\ &&
F_{UU,L} = \frac{x}{4z}F_0 (2\mathcal{S}_1^{\mu\nu})W_{U,\mu\nu} , 
\\  &&
F_{UU}^{\cos{\phi_h}} = \frac{x}{4z}F_0 (\mathcal{S}_3^{\mu\nu})W_{U,\mu\nu} , 
\\  &&
F_{UU}^{\cos{2\phi_h}} = \frac{x}{4z}F_0 (\mathcal{S}_0^{\mu\nu}-\mathcal{S}_1^{\mu\nu}-2\mathcal{S}_2^{\mu\nu})W_{U,\mu\nu} , 
\\  &&
F_{LU}^{\sin{\phi_h}} = \frac{x}{4z}F_0 (\mathcal{A}_3^{\mu\nu})W_{U,\mu\nu} ,
\\  &&
F_{UL}^{\sin{\phi_h}} = \frac{x}{4z}F_0 (\mathcal{S}_5^{\mu\nu})W_{L,\mu\nu} , 
\\  &&
F_{UL}^{\sin{2\phi_h}} = \frac{x}{4z}F_0 (-\mathcal{S}_4^{\mu\nu})W_{L,\mu\nu} , 
\\  &&
F_{LL} = \frac{x}{4z}F_0 (\mathcal{A}_1^{\mu\nu})W_{L,\mu\nu} , 
\\  &&
F_{LL}^{\cos{\phi_h}} = \frac{x}{4z}F_0 (\mathcal{A}_2^{\mu\nu})W_{L,\mu\nu} , 
\\  &&
F_{UT,T}^{\sin{(\phi_h-\phi_S)}} = \frac{x}{4z}F_0 (\mathcal{S}_1^{\mu\nu}-\mathcal{S}_0^{\mu\nu})W_{A,\mu\nu} ,
\\  &&
F_{UT,L}^{\sin{(\phi_h-\phi_S)}} = \frac{x}{4z}F_0 (2\mathcal{S}_1^{\mu\nu})W_{A,\mu\nu} , 
\\  &&
F_{UT}^{\sin{(\phi_h+\phi_S)}} = \frac{x}{4z}F_0\(-\frac{\mathcal{S}_4^{\mu\nu}}{2}W_{S,\mu\nu}-\frac{\mathcal{S}_0^{\mu\nu}-\mathcal{S}_1^{\mu\nu}-2\mathcal{S}_2^{\mu\nu}}{2}W_{A,\mu\nu}\) ,
\\  &&
F_{UT}^{\sin{(3\phi_h-\phi_S)}} = \frac{x}{4z}F_0\(-\frac{\mathcal{S}_4^{\mu\nu}}{2}W_{S,\mu\nu}+\frac{\mathcal{S}_0^{\mu\nu}-\mathcal{S}_1^{\mu\nu}-2\mathcal{S}_2^{\mu\nu}}{2}W_{A,\mu\nu}\),
\\  &&
F_{UT}^{\sin{\phi_S}} = \frac{x}{4z}F_0\(\frac{\mathcal{S}_5^{\mu\nu}}{2}W_{S,\mu\nu}-\frac{\mathcal{S}_3^{\mu\nu}}{2}W_{A,\mu\nu}\), 
\\  &&
F_{UT}^{\sin{(2\phi_h-\phi_S)}} = \frac{x}{4z}F_0\(\frac{\mathcal{S}_5^{\mu\nu}}{2}W_{S,\mu\nu}+\frac{\mathcal{S}_3^{\mu\nu}}{2}W_{A,\mu\nu}\), 
\\  &&
F_{LT}^{\cos{(\phi_h-\phi_S)}} = \frac{x}{4z}F_0 (\mathcal{A}_1^{\mu\nu})W_{S,\mu\nu} , 
\\  &&
F_{LT}^{\cos{\phi_S}} = \frac{x}{4z}F_0\(\frac{\mathcal{A}_2^{\mu\nu}}{2}W_{S,\mu\nu}+\frac{\mathcal{A}_3^{\mu\nu}}{2}W_{A,\mu\nu}\), 
\\ &&\label{def:FLT3p}
F_{LT}^{\cos{(2\phi_h-\phi_S)}} = \frac{x}{4z}F_0\(\frac{\mathcal{A}_2^{\mu\nu}}{2}W_{S,\mu\nu}-\frac{\mathcal{A}_3^{\mu\nu}}{2}W_{A,\mu\nu}\),
\end{eqnarray}
where
\begin{eqnarray}\label{def:F0}
F_0 = \frac{1}{\sqrt{1-\gamma^2\gamma_h^2-\gamma^2\frac{\bm{p}_\perp^2}{z^2Q^2}}\(1+\frac{\gamma^2}{2x}\)}.
\end{eqnarray}
The factor $F_0$ accounts for the overall kinematic factor that appears in the definition of the cross-section but is not separated from the definition of the structure functions. 

In the massless limit, $F_0=1$. However, when the computation of the hadron tensor is done assuming massless kinematics, the factor $F_0$ can still be taken exactly, since it originates from a mismatch of the definitions of the cross-section and the structure functions and is therefore not affected by any computation. In practice, $F_0$ is close to 1 for a typical data point. Nonetheless, for some HERMES bins \cite{HERMES:2012uyd} at $Q\sim 2.3$GeV, it can reach values $\sim 0.8$, making its inclusion in phenomenological analyses important.

The expressions (\ref{def:FUUT} - \ref{def:FLT3p}) are completely general and follow directly from SIDIS kinematics. They represent an alternative, although completely equivalent, formulation to those presented in refs.~\cite{Diehl:2005pc, Ebert:2021jhy}. We find our representation more practical, as it expresses the structure functions in terms of the kinematic vectors at hand, thus avoiding the introduction of photon helicity vectors. 

\section{Hadron tensor in TMD-with-KPCs factorization}
\label{subsec:hadron_tensor}

The TMD factorization theorem provides a general expression for the hadron tensor. The LP term has been derived in numerous previous works, see, e.g., refs \cite{Bacchetta:2006tn, Collins:2011zzd, Ebert:2021jhy}. Here, we generalize those derivations by including KPCs, as it is discussed in the introduction. Our approach employs the background field method \cite{Vladimirov:2021hdn, Balitsky:2025bup}, which, for this particular computation, closely resembles the Soft-Collinear Effective Field Theory (SCET) derivation \cite{Echevarria:2011epo, Ebert:2021jhy}. In a nutshell, the method involves splitting the fields into parts associated with the target and the produced hadrons, while integrating out the remaining components. In the strict limit $Q\to\infty$, our expression reproduces the LP result, but it is explicitly gauge- and frame-invariant. The theoretical foundation and a detailed discussion on the TMD-with-KPCs factorization theorem is given in ref.~\cite{Vladimirov:2023aot}.

The computation of the KPCs that follows the LP term consists in the extraction of the TMD-twist-two component from the higher power terms of the TMD factorization. This task is not as complicated as it sounds, due to the observation that the insertion of an additional gluon field into the operator necessarily increases its twist. Consequently, the computation of KPCs for the LP term effectively reduces to that for free quarks, analogous to the LP calculation itself. Let us mention that this observation is similar to the celebrated statement that collinear operators of twist-two correspond to those constructed from free quark fields \cite{Balitsky:1987bk}.

In ref.~\cite{Vladimirov:2023aot}, the KPCs following the LP term were derived to all orders for the Drell-Yan hadron tensor and subsequently summed up into a compact series. The resulting series is regular, allowing for a reordering of summations and a reformulation of the procedure into a more natural framework. Specifically, one can consider a prototype of TMD operator and then determine its TMD-twist-two component. Using these matrix elements as building blocks enables the calculation of the hadron tensor avoiding the tedious multipole expansion.

In this section, we present the explicit computation of the hadron tensor. The calculation is performed in the massless limit, since target-mass corrections for TMD factorization remain unknown. Hence, from this point onward, we set $M=m_h=0$. The useful expressions for this transition are collected in appendix \ref{app:massless}.

\subsection{Twist-two part of TMD correlators}

In the case of SIDIS, two types of correlators appear: those with hadrons in the initial state and those with hadrons in the final state. We denote them as
\begin{eqnarray}\label{def:Psi}
\widetilde{\Psi}^{\,[\Gamma]}(y)&=&\sum_{X}\Tr\langle P,S|\bar q(y)[y,y+n\infty]\frac{\Gamma}{2}|X\rangle \langle X|[+n\infty,0]q(0)|P,S\rangle,
\\\label{def:Theta}
\widetilde{\Theta}^{\,[\Gamma]}(y)&=&\frac{1}{N_c}\sum_X \Tr \langle 0|[-\bar n\infty +y,y]\frac{\Gamma}{2}q(y)|p_h,X\rangle\langle p_h,X|\bar q(0)[0,-\bar n\infty]|0\rangle,
\end{eqnarray}
where $\Gamma$ is a Dirac matrix, $[a,b]$ represents the gauge link connecting the points $a$ and $b$ and the trace operation $\Tr$ runs over color and spinor indices. The coordinate $y$ is fully general, i.e., it is not aligned to any light-cone direction. These correlators serve as prototypes of quark TMDPDFs and TMDFFs, respectively. Note that the associated Wilson lines in these operators point in opposite directions, as it appears in SIDIS factorization \cite{Collins:2002kn, Boer:2003cm}. In addition, we also need correlators with quark and anti-quark fields acting at different states
\begin{eqnarray}\label{def:Psi-bar}
\overline{\widetilde{\Psi}}^{\,[\Gamma]}(y)&=&\sum_X \Tr\langle P,S|[y+n\infty,y]q(y)|X\rangle \langle X|\bar q(y)[0,+n\infty]\frac{\Gamma}{2}|P,S\rangle,
\\\label{def:Theta-bar}
\overline{\widetilde{\Theta}}^{\,[\Gamma]}(y)&=&\frac{1}{N_c}\sum_X \Tr \langle 0|\bar q(y)[y,-\bar n\infty+y]|p_h,X\rangle\langle p_h,X|\frac{\Gamma}{2}[-\bar n\infty,0]q(0)|0\rangle,
\end{eqnarray}
which correspond to anti-quark distributions.

The correlators (\ref{def:Psi} -- \ref{def:Theta-bar}) do not possess a definite TMD-twist; rather, they are a sum of contributions with different twists. Meanwhile, the pure TMD-twist-two correlators are defined as
\begin{eqnarray}\label{def:Phi}
&&\widetilde{\Phi}_{11}^{\,[\Gamma^+]}(y)
\\\nn &&=\sum_{X}\Tr\langle P,S|\bar q(y^-n+y_T)[y^-n+y_T,y^+\bar n+y_T+n\infty]\frac{\Gamma^+}{2}|X\rangle \langle X|[+n\infty,0]q(0)|P,S\rangle,
\\\label{def:Delta}
&&\widetilde{\Delta}_{11}^{\,[\Gamma^-]}(y)
\\\nn &&=\frac{1}{N_c}\sum_X \Tr \langle 0|[-\bar n\infty +y^- n+y_T,y^+\bar n+y_T]\frac{\Gamma^-}{2}q(y^+\bar n+y_T)|p_h,X\rangle\langle p_h,X|\bar q(0)[0,-\bar n\infty]|0\rangle,
\end{eqnarray}
where $\Gamma^\pm\in\{\gamma^\pm,\gamma^\pm \gamma^5,i\sigma^{\alpha\pm}\gamma^5\}.$ These Dirac structures project out only the ``good'' components of quark spinors along the $n$ and $\bar n$ directions for $\Gamma^+$ and $\Gamma^-$, correspondingly. The definitions for the anti-quark correlators are analogous. It is worth noting that these matrix elements depend only on a single light-cone component of $y$, although we utilize the total vector $y$ as an argument.

The definite TMD-twist matrix elements represent irreducible non-perturbative functions, in the sense that they do not mix with any other non-perturbative functions under QCD evolution. This is their defining property, which follows from the symmetries of the corresponding operators \cite{Vladimirov:2021hdn, Rodini:2022wki}. In contrast, the generic TMD correlators (\ref{def:Psi} -- \ref{def:Theta-bar}) do not have specific evolution properties, and do not even obey a closed evolution equation, instead mixing with various multi-point correlation functions. Nonetheless, a generic correlator can be decomposed into irreducible components of increasing TMD-twists, which form a series of KPCs corresponding to particular twists. We remark that, although technically different, this procedure is conceptually close to the computation of KPCs in collinear factorization; see, for instance, ref.~\cite{Braun:2011dg}.

The theoretical justification for the TMD-twist-two extraction method is given in ref.~\cite{Vladimirov:2023aot}. Practically, the strategy is the following: to extract the TMD-twist-two component of the operators appearing in (\ref{def:Psi} -- \ref{def:Theta-bar}), one must expand the fields at coordinate $y$ along the corresponding light-cone direction ($\bar n$ for $\widetilde\Psi$, and $n$ for $\widetilde\Theta$). This generates a series of operators with growing derivatives, ($\partial_- q$, $\partial_- \bar q$) for $\widetilde\Psi$ and ($\partial_+ q$, $\partial_+ \bar q$) for $\widetilde\Theta$. By applying the quark equation of motion $\fnot D q=0$, these derivatives can be rewritten in terms of transverse and ``good'' derivatives of the quark fields
\begin{eqnarray}
\partial_\pm q=-\frac{\partial_T^2}{2\partial_\mp}q+...~,
\end{eqnarray}
where $1/\partial_\mp$ denotes the integration operator along the corresponding direction. The dots represent terms containing additional gluon fields, and thus contributing to higher-twist terms. Since the second quark field is independent on $y$, the $\partial_T$ and $\partial_\mp$ derivatives can be taken outside the operator. This yields
\begin{eqnarray}
\widetilde{\Psi}^{\,[\Gamma]}(y)&=&
\sum_{n=0}^\infty \frac{(-y^+)^n}{n!}\(\frac{\partial_T^2}{2\partial_+}\)^n\widetilde{\Psi}^{\,[\Gamma]}(y^-n+y_T)+...~,
\\
\widetilde{\Theta}^{\,[\Gamma]}(y)&=&
\sum_{n=0}^\infty \frac{(-y^-)^n}{n!}\(\frac{\partial_T^2}{2\partial_-}\)^n\widetilde{\Theta}^{\,[\Gamma]}(y^+\bar n+y_T)+...~,
\end{eqnarray}
with analogous expressions for the anti-quark correlators.

These formulas still have admixture of higher twists due to the presence of ``bad'' spinor components. Therefore, it is necessary to further decompose the spinors and express these remaining ``bad'' components in terms of the ``good'' ones by another application of the equations of motion. Then, the correlators read \cite{Vladimirov:2023aot}
\begin{eqnarray}\label{Psi:1}
\widetilde{\Psi}^{\,[\Gamma]}(y)&=&
\sum_{n=0}^\infty \frac{(-y^+)^n}{n!}\(\frac{\partial_T^2}{2\partial_+}\)^n
\Bigg[
\widetilde{\Phi}_{11}^{\,\llbracket\Gamma\rrbracket}(y^-n+y_T)
\\\nn &&
-\frac{\partial_{T\alpha}}{2\partial_+} \widetilde{\Phi}_{11}^{\,\llbracket \gamma^\alpha \gamma^+\Gamma+\Gamma \gamma^+\gamma^\alpha \rrbracket}(y^-n+y_T)
+\frac{\partial_{T\alpha}\partial_{T\beta}}{4\partial^2_+} \widetilde{\Phi}_{11}^{\,\llbracket \gamma^\alpha \gamma^+\Gamma \gamma^+\gamma^\beta \rrbracket}(y^-n+y_T)\Bigg]
+...~,
\\\label{Theta:1}
\widetilde{\Theta}^{\,[\Gamma]}(y)&=&
\sum_{n=0}^\infty \frac{(-y^-)^n}{n!}\(\frac{\partial_T^2}{2\partial_-}\)^n
\Bigg[
\widetilde{\Delta}_{11}^{\llbracket\Gamma\rrbracket}(y^+\bar n+y_T)
\\\nn &&
-\frac{\partial_{T\alpha}}{2\partial_-} \widetilde{\Delta}_{11}^{\,\llbracket \gamma^\alpha \gamma^-\Gamma+\Gamma \gamma^-\gamma^\alpha \rrbracket}(y^+\bar n+y_T)
+\frac{\partial_{T\alpha}\partial_{T\beta}}{4\partial^2_-} \widetilde{\Delta}_{11}^{\,\llbracket \gamma^\alpha \gamma^-\Gamma \gamma^-
\gamma^\beta \rrbracket}(y^+\bar n+y_T)\Bigg]
+...~,
\end{eqnarray}
where
\begin{eqnarray}\nn
\big\llbracket \Gamma \big\rrbracket &=&\[\frac{\gamma^+\gamma^-}{2}\Gamma\frac{\gamma^-\gamma^+}{2}\]\qquad \text{for~}\widetilde{\Psi},
\\\nn
\big\llbracket \Gamma \big\rrbracket &=&\[\frac{\gamma^-\gamma^+}{2}\Gamma\frac{\gamma^+\gamma^-}{2}\]\qquad \text{for~}\widetilde{\Theta}.
\end{eqnarray}
This latter form contains only TMD-twist-two distributions, while the higher-twist contributions are hidden in the dotted terms. Importantly, no other twist-two contributions appear in these correlators, since all eliminated terms contain at least one gluon field and thus have TMD-twist-three or higher.

The expressions (\ref{Psi:1}, {\ref{Theta:1}) are rather abstract because they incorporate a sum over powers of derivatives and integrals. This structure can be greatly simplified by moving to the momentum space representation. We define
\begin{eqnarray}\label{def:Phi-mom}
\Phi_{11}^{\,[\Gamma^+]}(\xi,k_T) &=&
\int \frac{dy^- d^2y_T}{(2\pi)^3} e^{-i\xi y^-P^+-i(yk)_T}\widetilde{\Phi}_{11}^{\,[\Gamma^+]}(y^-n+y_T),
\\\label{def:Delta-mom}
\Delta_{11}^{\,[\Gamma^-]}(\zeta,k_T)
&=&\frac{1}{2\zeta}\int \frac{dy^+ d^2y_T}{(2\pi)^3} e^{i\frac{1}{\zeta}y^+p_h^-+i(yk)_T}
\widetilde{\Delta}_{11}^{\,[\Gamma^-]}(y^+\bar n+y_T).
\end{eqnarray}
Here, the factor $1/2\zeta$ is part of the conventional TMDFF definition \cite{Bacchetta:2006tn, Metz:2016swz}. After the Fourier transformation, the derivatives act on the exponent, generating corresponding factors of momenta. The resulting series can then be explicitly computed, yielding
\begin{eqnarray}\label{Psi:2}
\widetilde{\Psi}^{\,[\Gamma]}(y)\Big|_{\text{tw2}}&=&
P^+\int d\xi d^2k_T e^{-i\frac{y^+k_T^2}{2\xi P^+}} e^{i(ky)_T+iy^-\xi P^+}
\Bigg[
\\\nn &&
\Phi_{11}^{\,\llbracket\Gamma\rrbracket}(\xi,k_T)
-\frac{k_{T\alpha}}{2\xi P^+} \Phi_{11}^{\,\llbracket \gamma^\alpha \gamma^+\Gamma+\Gamma \gamma^+\gamma^\alpha \rrbracket}(\xi ,k_T)
+\frac{k_{T\alpha}k_{T\beta}}{(2\xi P^+)^2} \Phi_{11}^{\,\llbracket \gamma^\alpha \gamma^+\Gamma \gamma^+\gamma^\beta \rrbracket}(\xi,k_T)\Bigg],
\\\label{Theta:2}
\widetilde{\Theta}^{\,[\Gamma]}(y)\Big|_{\text{tw2}}&=&
2\,p_h^- \int\frac{d\zeta d^2k_T}{\zeta}
e^{i\zeta\frac{y^-k_T^2}{2p_h^-}} e^{-i(ky)_T-iy^+p_h^-/\zeta}
\Bigg[
\\\nn &&
\Delta_{11}^{\,\llbracket\Gamma\rrbracket}(\zeta,k_T)
-\zeta\frac{k_{T\alpha}}{2p_h^-} \Delta_{11}^{\,\llbracket \gamma^\alpha \gamma^+\Gamma+\Gamma \gamma^+\gamma^\alpha \rrbracket}(\zeta,k_T)
+\zeta^2\frac{k_{T\alpha}k_{T\beta}}{(2p_h^-)^2} \Delta_{11}^{\,\llbracket \gamma^\beta \gamma^+\Gamma \gamma^+\gamma^\alpha \rrbracket}(\zeta,k_T)\Bigg].
\end{eqnarray}
The label $\text{tw2}$ indicates all higher-twist terms have been eliminated. These expressions can be simplified even further by observing that $(-k_T^2)/2k^\pm$ corresponds to $k^\mp$ if $k^2=0$. Thus, a more natural form to present the twist-two component of the TMD correlators is
\begin{eqnarray}\label{Psi:3}
\widetilde{\Psi}^{\,[\Gamma]}(y)\Big|_{\text{tw2}}&=&
P^+\int d^4k \delta(k^2) e^{i(yk)}\int d\xi \delta(k^+-\xi P^+)
\Bigg[
\\\nn &&
2k^+\Phi_{11}^{\,\llbracket\Gamma\rrbracket}(\xi,k_T)
-k_{T\alpha} \Phi_{11}^{\,\llbracket \gamma^\alpha \gamma^+\Gamma+\Gamma \gamma^+\gamma^\alpha \rrbracket}(\xi,k_T)
-k^-\frac{k_{T\alpha}k_{T\beta}}{k_T^2} \Phi_{11}^{\,\llbracket \gamma^\alpha \gamma^+\Gamma \gamma^+\gamma^\beta \rrbracket}(\xi,k_T)\Bigg],
\\\label{Theta:3}
\widetilde{\Theta}^{\,[\Gamma]}(y)\Big|_{\text{tw2}}&=&
2\,p_h^-\int d^4k \delta(k^2) e^{i(yk)}\int \frac{d\zeta}{\zeta}\delta\(k^--\frac{p_h^-}{\zeta}\)
\Bigg[
\\\nn &&
2k^-\Delta_{11}^{\,\llbracket\Gamma\rrbracket}(\zeta,k_T)
-k_{T\alpha} \Delta_{11}^{\,\llbracket \gamma^\alpha \gamma^-\Gamma+\Gamma \gamma^-\gamma^\alpha \rrbracket}(\zeta,k_T)
-k^+\frac{k_{T\alpha}k_{T\beta}}{k_T^2} \Delta_{11}^{\,\llbracket \gamma^\beta \gamma^-\Gamma \gamma^-\gamma^\alpha \rrbracket}(\zeta,k_T)\Bigg].
\end{eqnarray}
The anti-quark correlators satisfy analogous relations.

The last step is to consider all elements of the Dirac basis $\{1,\gamma^5,\gamma^\mu, \gamma^\mu\gamma^5,i\sigma^{\mu\nu}\gamma^5\}$ and to express each correlator explicitly in terms of the TMD distributions. The expressions for TMDPDFs are
\begin{eqnarray}\label{Psi:4}
\widetilde{\Psi}^{[1]}(y)\Big|_{\text{tw2}}&=&\widetilde{\Psi}^{[\gamma^5]}(y)\Big|_{\text{tw2}}=0,
\\
\widetilde{\Psi}^{[\gamma^\mu]}(y)\Big|_{\text{tw2}}&=&2P^+
\int d^4 k \,\delta(k^2)e^{i(ky)}\int d\xi \delta(k^+-\xi P^+)k^\mu \Phi_{11}^{[\gamma^+]}(\xi,k_T),
\\
\widetilde{\Psi}^{[\gamma^\mu\gamma^5]}(y)\Big|_{\text{tw2}}&=&2P^+
\int d^4 k \,\delta(k^2)e^{i(ky)}\int d\xi \delta(k^+-\xi P^+) k^\mu \Phi_{11}^{[\gamma^+\gamma^5]}(\xi,k_T),
\\
\widetilde{\Psi}^{[i\sigma^{\mu\nu}\gamma^5]}(y)\Big|_{\text{tw2}}&=&2P^+
\int d^4 k \,\delta(k^2)e^{i(ky)}\int d\xi \delta(k^+-\xi P^+) \Big(
\\\nn && 
g^{\mu\alpha}_T k^\nu-g^{\nu\alpha}_T k^\mu+\frac{k^\mu n^\nu-n^\mu k^\nu}{k^+}k_T^\alpha\Big)
\Phi_{11}^{[i\sigma^{\alpha +}\gamma^5]}(\xi,k_T).
\end{eqnarray}
These expressions were already derived in ref.~\cite{Vladimirov:2023aot}. The computation for the TMDFFs yields
\begin{eqnarray}
\widetilde{\Theta}^{[1]}(y)\Big|_{\text{tw2}}&=&\widetilde{\Theta}^{[\gamma^5]}(y)\Big|_{\text{tw2}}=0,
\\
\widetilde{\Theta}^{[\gamma^\mu]}(y)\Big|_{\text{tw2}}&=&4\,p_h^-
\int d^4 k \,\delta(k^2)e^{i(ky)}\int \frac{d\zeta}{\zeta} \delta\(k^--\frac{p^-_h}{\zeta}\)
k^\mu \Delta_{11}^{[\gamma^-]}(\zeta,k_T),
\\
\widetilde{\Theta}^{[\gamma^\mu\gamma^5]}(y)\Big|_{\text{tw2}}&=&4\,p_h^-
\int d^4 k \,\delta(k^2)e^{i(ky)}\int \frac{d\zeta}{\zeta} \delta\(k^--\frac{p^-_h}{\zeta}\)
k^\mu \Delta_{11}^{[\gamma^-\gamma^5]}(\zeta,k_T),
\\\label{Theta:sigma}
\widetilde{\Theta}^{[i\sigma^{\mu\nu}\gamma^5]}(y)\Big|_{\text{tw2}}&=&
4\,p_h^-
\int d^4 k \,\delta(k^2)e^{i(ky)}\int \frac{d\zeta}{\zeta} \delta\(k^--\frac{p^-_h}{\zeta}\)
\Big(
\\\nn && 
g^{\mu\alpha}_T k^\nu-g^{\nu\alpha}_T k^\mu+\frac{k^\mu \bar n^\nu-\bar n^\mu k^\nu}{k^-}k_T^\alpha\Big)
\Delta_{11}^{[i\sigma^{\alpha -}\gamma^5]}(\zeta,k_T).
\end{eqnarray}
The expressions for anti-quark correlators are analogous. We stress that these expressions are valid for any spin of the particle, including spinless and higher-spin cases.

In this form, it follows that the TMD-twist-two TMD correlators satisfy the set of equations
\begin{eqnarray}\label{diff_tw2}
&&\partial^2 \widetilde{\Psi}^{[\Gamma]}(y)\Big|_{\text{tw2}}=\partial^2 \widetilde{\Theta}^{[\Gamma]}(y)\Big|_{\text{tw2}}=0,
\\\nn
&& 
\partial_\mu \widetilde{\Psi}^{[\gamma^\mu]}(y)\Big|_{\text{tw2}}
=\partial_\mu \widetilde{\Psi}^{[\gamma^\mu\gamma^5]}(y)\Big|_{\text{tw2}}
=\partial_\mu \widetilde{\Psi}^{[i\sigma^{\mu\nu}\gamma^5]}(y)\Big|_{\text{tw2}}=0,
\\\nn
&& 
\partial_\mu \widetilde{\Theta}^{[\gamma^\mu]}(y)\Big|_{\text{tw2}}
=\partial_\mu \widetilde{\Theta}^{[\gamma^\mu\gamma^5]}(y)\Big|_{\text{tw2}}
=\partial_\mu \widetilde{\Theta}^{[i\sigma^{\mu\nu}\gamma^5]}(y)\Big|_{\text{tw2}}=0
\end{eqnarray}
which are also defining for the twist-two collinear correlators \cite{Balitsky:1987bk}. In the case of collinear twist, these equations can serve as a definition of the twist-two operators \cite{Geyer:1999uq}. Evidently, the same conclusion can be made for TMD-twist-two operators, i.e., one can define the TMD-twist-two component of a correlator as the part that satisfies the equations (\ref{diff_tw2}). This can be viewed as a manifestation of the Lorentz-invariance relations, and leads to an alternative derivation of the TMD-with-KPCs factorization theorem.

\subsection{TMD-with-KPC convolution for SIDIS}

The main ingredient required to construct the cross-section is the hadron tensor, which encapsulates the information associated with the strong interaction. In the case of SIDIS, it is given by
\begin{eqnarray}
W^{\mu\nu}= \int\frac{d^4y}{(2\pi)^4}e^{i(yq)}\sum_X \langle P|J^{\dagger \mu}(y)|p_h,X\rangle \langle X,p_h|J^\nu(0)|P\rangle,
\end{eqnarray}
where $J^\mu(y)$ denotes the electromagnetic (EM) current
\begin{eqnarray}
J^\mu(y) = e\,\bar{q}(y)\gamma^\mu q(y),
\end{eqnarray}
with $q(y)$ representing the quark field, and $e$ its electric charge. For shortness, we omit the flavor indices of the quarks. The sum over all active flavors is restored in sec.~\ref{subsec:StructFunctions}.

At tree order, the computation of the factorized expression reduces to the recoupling of spinor and color indices, which is carried out with the help of the Dirac decomposition. For any matrix $A$ in spinor space, one has
\begin{eqnarray}
A=\frac{1}{2}\sum_{a}\overline{\Gamma}_a A^{[\Gamma_a]},\qquad A^{[\Gamma]}=\frac{1}{2}\Tr\,(A \Gamma),
\end{eqnarray}
where $\Gamma_a\in\{1,\gamma^5,\gamma^\mu,\gamma^\mu\gamma^5,i\sigma^{\mu\nu}\gamma^5\}$, and 
$\overline{\Gamma}_a\in\{1,\gamma^5,\gamma^\mu,-\gamma^\mu\gamma^5,-i\sigma^{\mu\nu}\gamma^5/2\}$. Recoupling the indices yields
\begin{eqnarray}\label{eq:W-tree}
\nn
W^{\mu\nu}_{\text{tree}} &=& \frac{e^2}{4}\int \frac{d^4y}{(2\pi)^4} e^{i(yq)} \sum_{a,b}\bigg\{
\\ &&
\Tr(\gamma^{\mu}\overline{\Gamma}_b\gamma^\nu\overline{\Gamma}_a)
\widetilde{\Psi}^{[\Gamma_a]}(y)\widetilde{\Theta}^{[\Gamma_b]}(y)
+\Tr(\gamma^{\mu}\overline{\Gamma}_a\gamma^\nu\overline{\Gamma}_b)
\overline{\widetilde{\Psi}}^{[\Gamma_a]}(y)\overline{\widetilde{\Theta}}^{[\Gamma_b]}(y) \bigg\} + ...~.
\end{eqnarray}
Let us note that this expression is essentially identical to that obtained for the Drell-Yan process \cite{Piloneta:2024aac}, with the only differences being the interchange of the correlators $\widetilde{\Theta}\leftrightarrow \widetilde{\Psi}$ and the sign of the Fourier transform. 

The hadron tensor (\ref{eq:W-tree}) contains contributions of various twists, both in the explicit terms and in the dotted ones, which include the diagrams with radiation of additional gluons. As discussed in the previous section, any insertion of a gluon field (apart from those forming the Wilson lines) increases the TMD-twist of a term. Therefore, the dotted contribution consists solely of higher-twist operators, and does not contribute to the KPCs following the LP term. The TMD-twist-two part of the $\widetilde{\Psi}$ and $\widetilde{\Theta}$ correlators is presented in eqns.~(\ref{Psi:4}-\ref{Theta:sigma}).

Beyond tree level one needs to compute the contributions from hard fields, which can be done perturbatively. It has been proven \cite{Collins:2011zzd, Vladimirov:2023aot} that the only diagrams contributing to the LP term and it descendants are the virtual ones. The divergent part of the hard coefficient function renormalizes the TMD operators, inducing their scale dependence. Importantly, this procedure must be performed power by power in position space, and each power term has a distinct diagrammatic structure. Nevertheless, the resulting coefficient function is identical for all KPCs terms. This property follows from the QED Ward identity, which must be satisfied by the twist-two part of the hadron tensor independently. This has also been checked explicitly, see ref.~\cite{Vladimirov:2023aot}. 

Thus, after accounting for the perturbative corrections, we obtain a expression for the hadron tensor that incorporates all KPCs descendants of the LP term. We find
\begin{eqnarray}\label{eq:W-all}
\nn
W^{\mu\nu}_{\text{KPC}} &=& C_{0,\text{DIS}}\(\frac{Q^2}{\mu^2}\)\frac{e^2}{4}\int \frac{d^4y}{(2\pi)^4} e^{i(yq)} \sum_{a,b}\bigg\{
\Tr(\gamma^{\mu}\overline{\Gamma}_b\gamma^\nu\overline{\Gamma}_a)
\widetilde{\Psi}^{[\Gamma_a]}(y;\mu,\zeta)\Big|_{\text{tw2}}\widetilde{\Theta}^{[\Gamma_b]}(y;\mu,\bar \zeta)\Big|_{\text{tw2}}
\\ &&
+\Tr(\gamma^{\mu}\overline{\Gamma}_a\gamma^\nu\overline{\Gamma}_b)
\overline{\widetilde{\Psi}}^{[\Gamma_a]}(y;\mu,\zeta)\Big|_{\text{tw2}}\overline{\widetilde{\Theta}}^{[\Gamma_b]}(y;\mu,\bar \zeta)\Big|_{\text{tw2}} \bigg\}~,
\end{eqnarray}
where $C_{0,\text{DIS}}$ is the hard coefficient function for SIDIS. The scale $\mu$ denotes the hard factorization scale, within $\mu\sim Q$. The parameters $\zeta$ and $\bar \zeta$ are the rapidity factorization scales, which satisfy the relation $\zeta \bar \zeta=Q^4$.

The hard coefficient function differs from that in the Drell-Yan process by finite terms originating from the different imaginary parts of the corresponding amplitudes, $C_{0,\text{DIS}}=|c(Q^2)|^2$ versus $C_{0,\text{DY}}=|c(-Q^2)|^2$. The expression for $c$ is known up to four loops \cite{Lee:2022nhh}. Note that the argument of the coefficient function is $Q^2$ rather than the LP expression $2q^+q^-$. This is a consequence of the KPCs resummation, which restores de Lorentz invariant form of the result (see detailed discussion in ref.~\cite{Vladimirov:2023aot}).

The factorization of fields with different rapidities beyond the tree approximation modifies the KPCs structure. Namely, the transverse derivatives $\partial_\mu$ acting on the TMD operators are replaced by the ``long'' transverse derivatives $\partial_\mu+[\partial_\mu \mathcal{D}]\ln(\zeta/\bar \zeta)/2$ \cite{Rodini:2022wki, Rodini:2022wic, Jaarsma:2025ksf}. These corrections are crucial because they preserve the boost invariance of TMD factorization and prevent mixing between contributions of different TMD-twists. Simultaneously, they avoid the use of eqns. (\ref{Psi:2}, \ref{Theta:2}), since the summation of ``long derivatives'' is not a simple task. However, it still can be carried out at the specific normalization point
\begin{eqnarray}\label{def:zeta=zeta}
\zeta=\bar \zeta=Q^2.
\end{eqnarray}
At this stage, the additive term in the derivative vanishes, and the KPCs can be summed as discussed earlier. After the summation, the normalization point can be shifted using the evolution equations for the TMD distributions. Nevertheless, the resulting expressions become significantly more involved and include Fourier-convolution integrals. Therefore, from now on, we fix the rapidity scales at (\ref{def:zeta=zeta}), while the hard factorization scale $\mu$ remains unrestricted.

Finally, by substituting the expressions for the twist-two TMD correlators (\ref{Psi:4}-\ref{Theta:sigma}) into (\ref{eq:W-all}), we obtain the desired form of the hadron tensor within the TMD factorization theorem including all KPCs
\begin{eqnarray}\label{hadronTensor}
W^{\mu\nu}_{\text{KPC}}&=&8\,e^2\,P^+p_h^- C_{0,\text{DIS}}\(\frac{Q^2}{\mu^2}\)
\int d^4k_1 d^4k_2\delta(k_1^2)\delta(k_2^2) \delta^{(4)}(q+k_1-k_2)
\\\nn && \times
\int d\xi\frac{d\zeta}{\zeta}\delta(k_1^+-\xi P^+)\delta\(k_2^--\frac{p_h^-}{\zeta}\)\Bigg\{
\\ \nn&& -(k_1k_2)H_0^{\mu\nu} \(
\Phi_{11}^{[\gamma^+]}\Delta_{11}^{[\gamma^-]}+\overline{\Phi}_{11}^{[\gamma^+]}\overline{\Delta}_{11}^{[\gamma^-]}
+\Phi_{11}^{[\gamma^+\gamma^5]}\Delta_{11}^{[\gamma^-\gamma^5]}+\overline{\Phi}_{11}^{[\gamma^+\gamma^5]}\overline{\Delta}_{11}^{[\gamma^-\gamma^5]}\)
\\\nn &&
+i\epsilon^{\mu\nu\alpha\beta}k_{1\alpha}k_{2\beta}
\(
\Phi_{11}^{[\gamma^+]}\Delta_{11}^{[\gamma^-\gamma^5]}-\overline{\Phi}_{11}^{[\gamma^+]}\overline{\Delta}_{11}^{[\gamma^-\gamma^5]}
+\Phi_{11}^{[\gamma^+\gamma^5]}\Delta_{11}^{[\gamma^-]}-\overline{\Phi}_{11}^{[\gamma^+\gamma^5]}\overline{\Delta}_{11}^{[\gamma^-]}\)
\\\nn &&
+(k_1k_2)\Bigg[
H_0^{\mu\nu}\(g^{\alpha\beta}+\frac{k_1^\alpha k_2^\beta}{k_1^+k_2^-}
-\frac{(k_1^\alpha k_2^+-k_2^\alpha k_1^+)(k_2^\beta k_1^--k_1^\beta k_2^-)}{(k_1k_2)k_1^+k_2^-}\)
\\\nn &&\qquad
-H_0^{\mu\alpha}H_0^{\nu\beta}-H_0^{\mu\beta}H_0^{\nu\alpha}
+\frac{N^\mu H_0^{\nu \beta}k_1^\alpha + N^\nu H_0^{\mu \beta}k_1^\alpha}{k_1^+}
+\frac{\bar N^\mu H_0^{\nu \alpha}k_2^\beta + \bar N^\nu H_0^{\mu \alpha}k_2^\beta}{k_2^-}
\\\nn &&\qquad
-\frac{(N^\mu\bar N^\nu +\bar N^\mu N^\nu)k_1^\alpha k_2^\beta}{k_1^+ k_2^-}
\Bigg]
\Big(
\Phi_{11}^{[i\sigma^{\alpha+}\gamma^5]}\Delta_{11}^{[i\sigma^{\beta-}\gamma^5]}+\overline{\Phi}_{11}^{[i\sigma^{\alpha+}\gamma^5]}\overline{\Delta}_{11}^{[i\sigma^{\beta-}\gamma^5]}\Big)\Bigg\},
\end{eqnarray}
where the arguments of $\Phi$ and $\Delta$ are $(\xi,k_{1T};\mu,Q^2)$ and $(\zeta,k_{2T};\mu,Q^2)$, respectively, and 
\begin{eqnarray}
H^{\mu\nu}_0=g^{\mu\nu}-\frac{k_1^\mu k_2^\nu+k_2^\mu k_1^\nu}{(k_1k_2)},\qquad
N^\mu =H_0^{\mu+},\qquad \bar N^\mu =H_0^{\mu-}.
\end{eqnarray}
It is worth noting that these tensors are defined such that
\begin{eqnarray}\label{transverse-1}
k_{1\mu}H_0^{\mu\nu}=k_{2\mu}H_0^{\mu\nu}=0,\qquad\text{if~}k_1^2=k_2^2=0.    
\end{eqnarray}

The expression (\ref{hadronTensor}) constitutes one of the main results of this paper. It represents the complete twist-two part of the SIDIS hadron tensor. Therefore, it can be considered as an extension of the TMD factorization, which we refer to as the TMD-with-KPCs factorization theorem.

The TMD-with-KPCs factorized expression incorporates the LP term (see below) as well as power-suppressed terms such that together form a gauge- and frame-invariant object. The correction terms to this factorization include two types of contributions (see also fig.~\ref{fig:power_pyramid}). First, there are higher-twist terms starting from the tw2$\times$tw3 term \cite{Rodini:2023plb}, which is accompanied by an extra suppressing factor of $1/Q$. Second, there are $q_T/Q$ terms originating from the contribution of tw2$\times$tw3 singular at $y_T^2\to0$. This singularity can be isolated and rewritten in terms of tw2$\times$tw2 terms accompanied by a $q_T/Q$ factor, thus representing a $q_T/Q$ correction \cite{Arroyo-Castro:2025slx}. Each of these corrections generates its own series of KPCs, which could be potentially summed. In this way, the TMD-with-KPCs factorization theorem is a systematic framework that expands the hadron tensor in powers of $\Lambda/Q$ and $q_T/Q$ while exactly accounts for $k_T/Q$ corrections.

The expression (\ref{hadronTensor}) is transverse with respect to $q_\mu$ 
\begin{eqnarray}
q_{\mu}W_{\text{KPC}}^{\mu\nu}=0,
\end{eqnarray}
as it follows from eqn.~(\ref{transverse-1}). Furthermore, we observe that
\begin{eqnarray}
k_{1\mu}W_{\text{KPC}}^{\mu\nu}=k_{2\mu}W_{\text{KPC}}^{\mu\nu}=0.
\end{eqnarray}
These conditions do not arise from the requirement of charge conservation, but rather are a consequence of the twist-two approximation (\ref{diff_tw2}). Furthermore, it can be shown \cite{Vladimirov:2023aot} that accounting for the KPCs also restores reparameterization-invariance \cite{Manohar:2002fd}, and thus the TMD-with-KPCs approach is frame-invariant.

A characteristic feature of the TMD-with-KPCs factorization theorem is the convolution integral appearing in (\ref{hadronTensor}), which is Lorentz invariant and represents the phase-space integration for the partonic scattering $q(k_1)+\gamma(q)\to q(k_2)$ involving exactly massless, non-collinear partons. The hadron momenta specify the collinear and transverse directions, thereby determining the momentum fractions and the transverse momenta of the partons. This can be interpreted as a Lorentz-invariant way of representing the partonic sub-process in its exact center-of-mass frame (i.e., accounting for the partons' transverse momentum) where the factorization theorem is derived.

To better understand the structure of the convolution integral, let us consider the phase-space domain of the partons. Taking into account the 8 delta-functions presented in (\ref{hadronTensor}), the integration can be reduced to a two-dimensional integral. A physically motivated choice of independent variables is $\{\vec k_1^2, \vec k_2^2\}$. The triangle inequality for the vector sum $\vec k_1+\vec q_T=\vec k_2$ restricts their absolute values to the region $R_T$, shown as the orange domain in fig.~\ref{fig:integration_domain}. This is also the domain of integration for the LP expression.

Using $\vec k_{1,2}^2$ as independent variables, the parton momentum fractions take the form
\begin{eqnarray}\label{eq:momentum_fractions}
\xi = \frac{x_1}{2}\(1 - \frac{\vec k_1^2}{\tau^2} + \frac{\vec k_2^2}{\tau^2}+\frac{\sqrt{\lambda(\vec k_1^2, \vec k_2^2, -\tau^2)}}{\tau^2}\), 
\quad 
\zeta = \frac{z}{2}\(1 - \frac{\vec k_1^2}{\vec k_2^2} - \frac{\tau^2}{\vec k_2^2}+\frac{\sqrt{\lambda(\vec k_1^2, \vec k_2^2, -\tau^2)}}{\vec k_2^2}\),
\end{eqnarray}
where $\lambda$ is the triangle function,
$$
\lambda(a,b,c)=a^2+b^2+c^2-2ab-2ac-2bc.
$$
This implies that the collinear momentum fractions of the partons are not fixed, as in the LP factorization, but instead vary depending on the kinematics. The causality requires $0<\xi<1$ and $0<\zeta<1$, which restricts the integration region for $\vec k_{1,2}^2$ to $R_{\xi \zeta}$, which is shown as the blue domain in fig.~\ref{fig:integration_domain}. 

\begin{figure}
\centering
\includegraphics[width=0.90\linewidth]{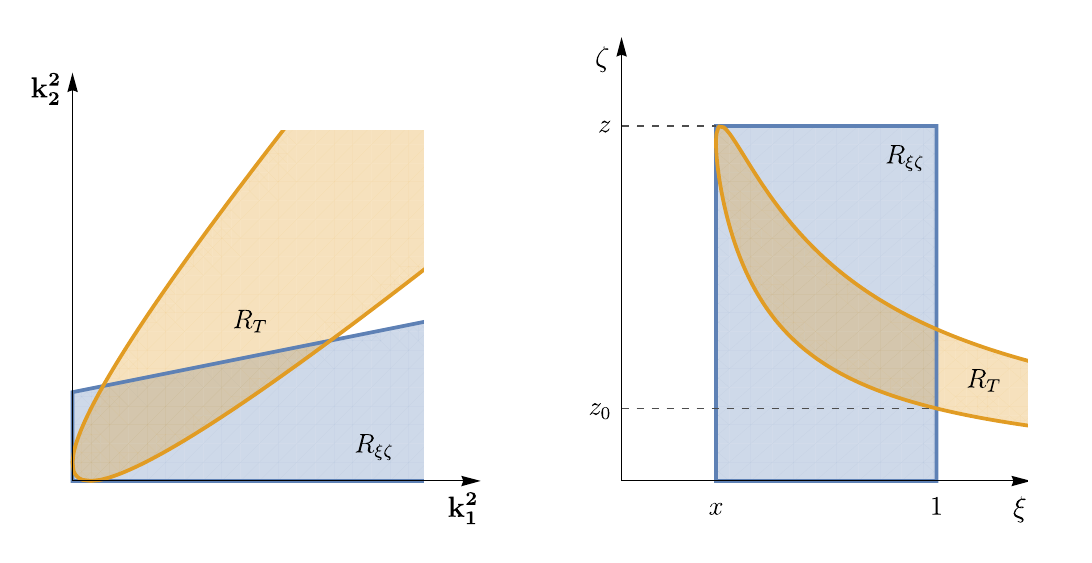}
\caption{Integration domain for the convolution integral (\ref{hadronTensor}) in $\{\vec k_1^2,\vec k_2^2\}$(left) and $\{\xi,\zeta\}$(right) planes. The region $R_T$ is generated by the constraint $\vec k_1+\vec q_T=\vec k_2$, while the region $R_{\xi\zeta}$ is defined by the condition $0<\xi,\zeta<1$.}
\label{fig:integration_domain}
\end{figure}

Together, the regions $R_T$ and $R_{\xi\zeta}$ form a finite domain of integration (see fig.~\ref{fig:integration_domain}). This is in contrast to the LP convolution, which extends to infinity. Although this may seem unusual, such restrictions are physically appropriate, since they prevent a parton from carrying momentum larger than that allowed by the external kinematics. Simultaneously, the collinear momentum fractions are not exact, but corrected for transverse momentum effects, and within the convolution integral, they cover the domain illustrated in the right panel of fig.~\ref{fig:integration_domain}. Note that the effective $\zeta$ of the fragmentation function is lower than the observed $z$, while the effective $\xi$ is higher than the observed $x$. The figure also marks the lower-limit point $z_0$, which is given by 
\begin{eqnarray}
z_0 = \frac{xz\tau^4}{(\tau^2+x\vec q_T^2)^2}\bigg[1+(2-x)\frac{\vec q_T^2}{\tau^2}-2\sqrt{(1-x)\frac{\vec q_T^2}{\tau^2}\left(1+\frac{\vec q_T^2}{\tau^2}\right)}\,\bigg].
\end{eqnarray}
Practically, the computation of the integral convolution is rather cumbersome, but it can be significantly simplified with a suitable change of variables. We present the details of the implementation of this convolution within the \texttt{artemide} code \cite{artemide} in the appendix \ref{app:convolution_integral}.

To obtain the LP limit, one must assume that $xP^+\gg 0$ and $p_h^-/z\gg 0$, while $\vec k_{1,2}^2$ are held fixed. This limit corresponds to the strict case $Q^2\to\infty$, with $\vec q_T^2$ also fixed, taken in the Breit frame. Note that this limit must be taken under the sign of the integral, which is not mathematically rigorous, since the integration region is restricted by $\sim 1/Q$. To guarantee the existence of the limit in this form, one should assume that the TMD distributions decay sufficiently fast at $\vec k^2\to\infty$. Then, the functions for $\xi$ and $\zeta$ (\ref{eq:momentum_fractions}) reduce to their LP counterparts with $x_{\text{LP}}=-q^+/P^+$ and $z_{\text{LP}}=p_h^-/q^-$, respectively. The convolution integral turns into the LP one (for the proportionality coefficient see (\ref{eq:convolution-LP})) setting the hierarchy for partons momenta as $k_1^+\gg k_{1T}\gg k_1^-$ and $k_2^-\gg k_{2T}\gg k_2^+$. The components of hadron tensor (\ref{hadronTensor}) in this limit are
\begin{eqnarray}
\lim_{\text{LP}}H_0^{\mu\nu}=g_T^{\mu\nu},\qquad
\lim_{\text{LP}}N^{\mu}=n^{\mu},\qquad
\lim_{\text{LP}}\bar N^{\mu}=\bar n^{\mu}.
\end{eqnarray}
Combining these elements yields the well-known expression for the SIDIS hadron tensor at LP \cite{Mulders:1995dh}.

\subsection{SIDIS structure functions}
\label{subsec:StructFunctions}

To obtain the explicit expressions for the structure functions, one must substitute the parameterizations of the TMD distributions. The TMDPDFs for a spin-1/2 polarized hadron are defined as
\begin{eqnarray}\label{def:psi+}
\Phi^{[\gamma^+]}(\xi,\vec k)&=&f_1(\xi,\vec k)-\frac{\epsilon^{\alpha\beta}_T k_{\alpha}s_{T\beta}}{M}f_{1T}^{\perp}(\xi,\vec k), 
\\
\Phi^{[\gamma^+\gamma^5]}(\xi,\vec k)&=&\lambda g_1(\xi,\vec k)-\frac{(k\cdot s_T)}{M}g_{1T}^\perp(\xi,\vec k),
\\
\Phi^{[i\sigma^{\alpha +}\gamma^5]}(\xi,\vec k)&=&s_T^{\alpha}h_1(\xi,\vec k)+\lambda\frac{\vec k^\alpha}{M}h_{1L}^\perp(\xi,\vec k)
\\\nn
&&-\frac{2\vec k^\alpha \vec k^\beta-\vec k^2g_T^{\alpha\beta}}{2M^2}s_{T\beta}h_{1T}^\perp(\xi,\vec k)- \frac{\epsilon_T^{\alpha\beta}\vec k_{\beta}}{M}h_1^\perp(\xi,\vec k).
\end{eqnarray}
The corresponding anti-quark distributions are defined analogously, except that the anti-quark functions $g_1$ and $g_{1T}^\perp$ acquire an additional factor of (-1) \cite{Mulders:1995dh}. 

The TMDFFs for a spin-0 hadron are
\begin{eqnarray}
\Delta^{[\gamma^-]}(\zeta,\vec k)&=&D_1(\zeta,\vec k), 
\\
\Delta^{[\gamma^-\gamma^5]}(\zeta,\vec k)&=&0,
\\\label{def:delta-sigma}
\Delta^{[i\sigma^{\alpha -}\gamma^5]}(\zeta,\vec k)&=&-\frac{\epsilon_T^{\alpha\beta}\vec k_{\beta}}{M}H_1^\perp(\zeta,\vec k).
\end{eqnarray}
It should be noted that the TMDFF $H_1^\perp$ is defined with a relative minus sign with respect to $h_1^\perp$ \cite{Bacchetta:2006tn, Rodini:2023plb}.

By inserting the parameterizations (\ref{def:psi+} - \ref{def:delta-sigma}) into eqn.~(\ref{hadronTensor}) and contracting with the lepton tensor components (\ref{def:S0} -- \ref{def:A3}) as given by the pocket formulas (\ref{def:FUUT} -- \ref{def:FLT3p}), we obtain the complete set of SIDIS structure functions. This computation is rather tedious, since the lepton tensor is naturally defined in the laboratory frame, while the hadron tensor is defined in the partonic center-of-mass frame. Consequently, their convolution produces a large number of terms that lack a direct physical or geometrical interpretation, representing instead projections of tensors from one frame onto another. The contributions proportional to the chiral-odd TMD distributions $\Phi^{[i\sigma^{\alpha+}\gamma^5]}$ and $\Delta^{[i\sigma^{\alpha-}\gamma^5]}$ are particularly cumbersome, as they involve tensors of higher rank.

To present the structure functions in a compact manner, we introduce the following short-hand notation for the TMD-with-KPCs convolution
\begin{eqnarray}\label{eq:convolution}
\mathcal{C}[A,f_1f_2] &=& 
F_0Q^4
\sum_{f} e_f^2 ~C_{0,\text{DIS}}\(\frac{Q^2}{\mu^2}\) 
\\\nn &&
\times \int d^4k_1 d^4k_2 \delta^{(4)}(q+k_1-k_2) \delta(k_1^2)\delta(k_2^2)
\int d\xi \frac{d\zeta}{\zeta} \delta\(k_1^+-\xi P^+\)\delta\(k_2^--\frac{p_h^-}{\zeta}\)
\\ \nn &&
\times (f_{1,f}(\xi,\vec k_1)f_{2,f}(\zeta,\vec k_2)+\overline{f}_{1,f}(\xi,\vec k_1)\overline{f}_{2,f}(\zeta,\vec k_2))A,
\end{eqnarray}
where the factor $F_0$ is defined in eqn.~(\ref{def:F0}), $f$ labels the active quark flavors, and $A$ denotes an expression constructed from the kinematic variables. The unpolarized structure functions then read
\begin{eqnarray}
F_{UU,T}&=&\mathcal{C}[1,f_1D_1]+\frac{1}{2} F_{UU,L},
\\
F_{UU,L}&=&
\mathcal{C}[4a(1+a),f_1D_1]
+
\mathcal{C}[-4a(1+a)\frac{(\vec k_1\vec k_2)}{m_hM},h^\perp_1H^\perp_1]
,
\\
F_{UU}^{\cos\phi_h}&=&
\mathcal{C}\[
2(1+2a)\frac{(kh)_\perp}{Q}
,f_1D_1\]
\\\nn &&
+
\mathcal{C}\[\frac{-2a(1+a)Q}{m_h M}\Big\{
(1+b)\(2(kh)_\perp+\sqrt{\vec q_T^2}\)-(\vec k_1\vec h_T)\(1+\frac{\vec q_T^2}{Q^2}\)\Big\}
,h^\perp_1H^\perp_1\]
,
\\
F_{UU}^{\cos2\phi_h}&=&-\frac{1}{2}F_{UU,L}+
\mathcal{C}\[
\frac{4(kh)_\perp^2}{Q^2}
,f_1D_1\]
\\\nn &&
+
\mathcal{C}\Big[
\frac{1}{m_h M}\Big\{
\((\vec k_1\vec k_2)-2 (\vec k_1\vec h_T)(\vec k_2\vec h_T)\)\(1+2\frac{(\vec k_1\vec k_2)}{Q^2}\)
\\\nn &&\qquad
+2\frac{\vec k_1^2\vec k_2^2}{Q^2}-4a(1+a)(\vec k_1\vec k_2)
\Big\}
,
h^\perp_1H^\perp_1\Big].
\end{eqnarray}
The longitudinal structure functions are
\begin{eqnarray}
F_{LU}^{\sin\phi_h}&=&0,
\\
F_{UL}^{\sin\phi_h}&=&\mathcal{C}\[\frac{-2a(1+a)Q}{m_h M}\Big\{
(\vec k_1\vec h_T)+(\vec k_2\vec h_T)-(kh)_\perp \(1-\frac{\vec q_T^2}{Q^2}\)
\Big\},h^\perp_{1L}H_{1}^\perp\],
\\
F_{UL}^{\sin2\phi_h}&=&\mathcal{C}\Bigg[\frac{1}{m_hM}\Big\{
((\vec k_1\vec k_2)-2(\vec k_1\vec h_T)(\vec k_2\vec h_T))
(1+a+b)
\\\nn &&\qquad
-a(1+a)(\vec k_1\vec q_T)\(b+\frac{\vec q_T^2}{Q^2}\)
+a(2+a)(1+b)(\vec k_2\vec q_T)
\Big\},h^\perp_{1L}H_{1}^\perp\Bigg],
\\
F_{LL}&=&\mathcal{C}\[1+2a,g_{1L}D_{1}\],
\\
F_{LL}^{\cos\phi_h}&=&\mathcal{C}\[2\frac{(kh)_\perp}{Q},g_{1L}D_{1}\].
\end{eqnarray}
Finally, the transverse structure functions take the form
\begin{eqnarray}
F_{UT,T}^{\sin(\phi_h-\phi_S)}&=&\mathcal{C}\[
\frac{-(\vec k_1 \vec h_T)}{M}
,f^\perp_{1T}D_{1}\]+\frac{1}{2}F_{UT,L}^{\sin(\phi_h-\phi_S)},
\\
F_{UT,L}^{\sin(\phi_h-\phi_S)}&=&
\mathcal{C}\[\frac{4a(1+a)(\vec k_1 \vec h_T)}{M},f^\perp_{1T}D_{1}\]
+\mathcal{C}\[2a(1+a)\frac{\vec k^2_1(\vec k_2 \vec h_T)}{m_hM^2},h^\perp_{1T}H^\perp_{1}\]
\\\nn &&
+\mathcal{C}\[-\frac{4}{m_hM}\frac{1+a}{1+b}\frac{(\vec k_1\vec h_T)\vec k_2^2-(\vec k_1\vec k_2)\sqrt{\vec q_T^2}}{Q^2},h^\perp_{1}H^\perp_{1}\],
\\
F_{UT}^{\sin(\phi_h+\phi_S)}&=&\mathcal{C}\[\frac{1}{M}\Big\{2\frac{(kh)_\perp}{Q^2}\,(a(\vec k_1 \vec q_T)-\vec k_1^2)+a(1+a)(\vec k_1 \vec h_T)\Big\},f_{1T}^\perp D_1\]
\\\nn && 
+ \mathcal{C}\Big[\frac{1}{m_h}\frac{1+a}{1+b}\Bigg\{(kh)_\perp\Big((1+b)^2+(4+2a+b-ab)\frac{\vec q_T^2}{Q^2}-a\frac{\vec q_T^4}{Q^4}\Big)
\\\nn &&
+(1-b)(1+a)^2\sqrt{\vec q_T^2}\Bigg\},h^\perp_{1}H_1^\perp\Big]
\\\nn && \qquad
+ \mathcal{C}\[-\frac{a\vec k_1^2}{4m_hM^2}\(2b(kh)_\perp-a(2+a+3b)\sqrt{\vec q_T^2}+a^2 \frac{\vec q_T^{3/2}}{Q^2}\),h_{1T}^\perp H_1^\perp\],
\\
F_{UT}^{\sin(3\phi_h-\phi_S)}&=&
\mathcal{C}\Big[\frac{1}{M_1}\Bigg\{
4\frac{(\vec k_1\vec h_T)^2(kh)_\perp}{Q^2}-3a(1+b)(kh)_\perp
\\\nn &&\qquad
-\frac{a^2\sqrt{\vec q_T^2}}{2}
\Big(2+a+b-a\frac{\vec q_T^2}{Q^2}\Big)
\Bigg\},f_{1T}^\perp D_1\Big]
\\\nn &&
+ \mathcal{C}\Bigg[-\frac{1}{2(1+b)m_h}\Bigg\{8b\frac{(kh)_\perp(\vec k_1 \vec h_T)^2}{Q^2}
-2a(\vec k_1 \vec h_T)(b(1+b)-a)
\\\nn &&\qquad
-2ab(3+2b)(kh)_\perp
+ a^2(2+b(b-a))\sqrt{\vec q_T^2}+\frac{a^3b}{Q^2}\vec q_T^3\Bigg\},h^\perp_{1}H_1^\perp\Bigg]
\\\nn && 
+ \mathcal{C}\Bigg[-\frac{(1+a)(1+b)}{2m_hM^2}\Bigg\{(kh)_\perp\big(3\vec k_1^2-4(\vec k_1 \vec h_T)^2-(2-a)a\vec q_T^2\big)
\\\nn &&\qquad
+ \sqrt{\vec q_T^2}\bigg((1+a)\vec k_1^2+(1-a)a^2\vec q_T^2-2(\vec k_1 \vec h_T)^2\bigg)\Bigg\},h_{1T}^\perp H_1^\perp\Bigg],
\\
F_{UT}^{\sin\phi_S}&=&\mathcal{C}\[\frac{(1+2a)}{M}\left\{a\frac{(\vec k_1 \vec h_T)\sqrt{\vec q_T^2}}{Q}-\frac{\vec k_1^2}{Q}\right\},f_{1T}^\perp D_{1}\] 
\\\nn && 
+ \mathcal{C}\[\frac{1}{m_hQ}\frac{(1+a)}{(1+b)}\Bigg\{2(\vec k_1 \vec k_2)\left((1+b)+\frac{\vec q_T^2}{Q^2}\right)-\frac{2(\vec k_1 \vec h_T)\vec k_2^2\sqrt{\vec q_T^2}}{Q^2}\Bigg\},h^\perp_{1}H_1^\perp\] 
\\\nn &&
+ \mathcal{C}\[-\frac{a}{m_hM^2}\left\{\frac{\vec k_1^2}{Q}((1+a)(\vec k_1 \vec k_2)-a\vec k_2^2)\right\},h_{1T}^\perp H_1^\perp\],
\\
F_{UT}^{\sin(2\phi_h-\phi_S)}&=&\mathcal{C}\[-\frac{(1+2a)}{M}\left\{\frac{\vec k_1^2}{Q}+a\frac{(\vec k_1 \vec h_T)\sqrt{\vec q_T^2}}{Q}-2\frac{(\vec k_1 \vec h_T)^2}{Q}\right\},f_{1T}^\perp D_{1}\]
\\\nn &&
+ \mathcal{C}\[-\frac{2}{m_2Q}\frac{(1+a)}{(1+b)}\left\{2b(\vec k_1 \vec h_T)^2-\vec k_1^2\left(\frac{\sqrt{\vec q_T^2}}{Q^2}(\vec k_1 \vec h_T)+b\right)\right\},h^\perp_{1}H_1^\perp\]
\\\nn &&
+\mathcal{C}\[\frac{(1+a)}{m_hM^2}\frac{\vec k_1^2}{Q}\left\{2(\vec k_1 \vec h_T)^2+(1-a)(\vec k_1 \vec q_T)\sqrt{\vec q_T^2}-(\vec k_1 \vec k_2)+( kh)_\perp\sqrt{\vec q_T^2}\right\},h_{1T}^\perp H_1^\perp\],
\\
F_{LT}^{\cos(\phi_h-\phi_S)}&=&\mathcal{C}\[
-(1+2a)\frac{(\vec k_1 \vec h_T)}{M}
,g_{1T}^\perp D_{1}\],
\\
F_{LT}^{\cos\phi_S}&=&\mathcal{C}\[\frac{1}{M}\left\{a\frac{(\vec k_1 \vec h_T)\sqrt{\vec q_T^2}}{Q}-\frac{\vec k_1^2}{Q}\right\},g_{1T}^\perp D_{1}\],
\\
F_{LT}^{\cos(2\phi_h-\phi_S)}&=&\mathcal{C}\[\frac{1}{M}\left\{\frac{\vec k_1^2}{Q}+a\frac{(\vec k_1 \vec h_T)\sqrt{\vec q_T^2}}{Q}-2\frac{(\vec k_1 \vec h_T)^2}{Q}\right\},g_{1T}^\perp D_{1}\].
\end{eqnarray}
In these expressions 
\begin{eqnarray}
a=\frac{z}{\zeta}-1=\frac{x}{\xi}\frac{\vec k_1^2}{Q^2},\qquad b=\frac{\xi}{x}-1=\frac{\zeta}{z}\frac{\vec k_2^2}{Q^2}-\frac{\vec q_T^2}{Q^2},
\end{eqnarray}
\begin{eqnarray}
(\vec k_{1,2} \vec h_T)=\frac{(\vec k_{1,2} \vec q_T)}{\sqrt{\vec q_T^2}},\qquad (kh)_\perp=\frac{(kp)_\perp}{\sqrt{\vec p_\perp^2}},\qquad (kp)_\perp=z\frac{(b-a)Q^2-a\vec q_T^2}{2}.
\end{eqnarray}
Let us remark that the variables $a$ and $b$ behave as $\mathcal{O}(Q^{-2})$. On the other hand, the transverse scalar products can be formulated as
\begin{eqnarray}
(\vec k_1\vec k_2)=\frac{\vec k_1^2+\vec k_2^2-\vec q_T^2}{2},
\qquad
(\vec k_1\vec q_T)=\frac{\vec k_2^2-\vec k_1^2-\vec q_T^2}{2},
\qquad
(\vec k_2\vec q_T)=\frac{\vec k_2^2-\vec k_1^2+\vec q_T^2}{2}.
\end{eqnarray}
The integrands are written in a form that makes their regularity in the limit $Q\to\infty$ explicit. 

These expressions constitute the main result of this work. In the limit $Q\to\infty$, they can be compared with the known results at LP, NLP and partially NNLP \cite{Bacchetta:2006tn, Bacchetta:2008xw, Boer:2011xd, Ebert:2021jhy, Rodini:2022wic}, and they are found to be in agreement with them. The corresponding expressions in this limit are presented in appendix \ref{app:largeQ}.

\section{Size estimation of KPCs}
\label{sec:impact_studies}

\begin{figure}
\begin{center}
\includegraphics[width=0.75\textwidth]{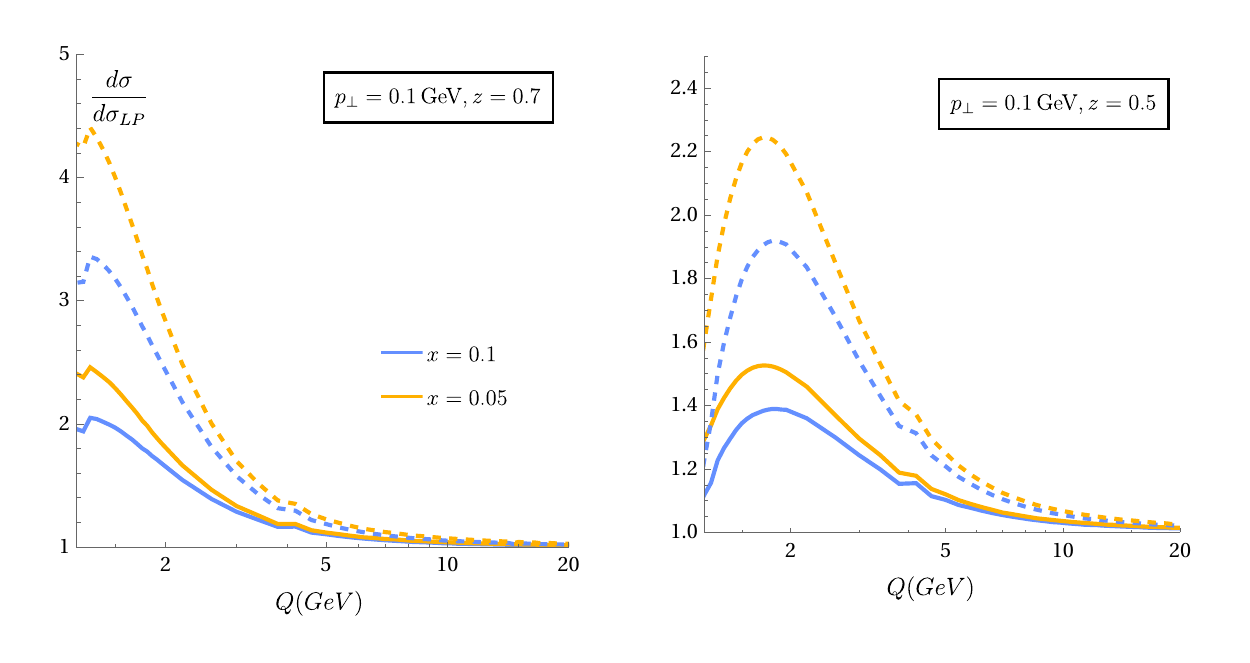}
\caption{\label{fig:1} Comparison of TMD-with-KPCs prediction to the pure LP TMD factorization expression for angle-integrated SIDIS cross-section as a function of $Q$. Solid lines indicate the comparison of $F_{UU,T}$ only, while dashed lines include the contribution $+\varepsilon F_{UU,L}$ term. The values of $(x,z,p_\perp)$ are indicated in the figure. The value of $\varepsilon$ is defined by (\ref{def:varepsilon}) and is $\varepsilon\sim 0.85-1.$ for the present kinematics.}
\end{center}
\end{figure}

In this section, we examine the potential phenomenological impact of including KPCs in SIDIS observables. For the numerical estimations, we use the \texttt{artemide} code \cite{Scimemi:2019cmh} with unpolarized TMD distributions (both TMDPDFs and TMDFFs) extracted in the ART25 fit \cite{Moos:2025sal}. This extraction incorporates more than a thousand points from Drell-Yan and SIDIS reactions measured over a wide range of $Q$ values, from $2$GeV in SIDIS up to 1TeV for neutral boson production in Drell-Yan. This analysis relies on the most advanced theoretical setup, including perturbative ingredients up to three and four-loop order (reaching the overall N$^4$LL perturbative accuracy), as well as other elements, such as large-$x$ resummation \cite{delRio:2025qgz}, and transverse momentum moments \cite{delRio:2024vvq}. For a complete description of ART25, we refer to the original publication.

The ART25 determination is performed using the LP TMD factorization theorem. Consequently, the values of the extracted TMD distributions are biased toward this order. In some cases, this bias may be negligible, and the extractions based on the LP TMD and TMD-with-KPCs formalisms would result into the same TMD distributions. For instance, in the Drell-Yan process, most measurements are made at large $Q$, with the most precise data obtained at Z-boson mass, $Q\sim 90$GeV. In this regime, power corrections are small and have little impact on the determination of TMDPDFs \cite{Piloneta:2024aac} (note, however, that a significant portion of the Drell-Yan data is taken at $Q\sim 8-16$GeV, where the KPCs effects become relevant). In contrast, the situation in SIDIS is exactly the opposite, as we demonstrate below. For that reason, the values of the TMDFFs are concerned and we could not compare with the data. A proper comparison would require a dedicated TMDFFs extraction within the updated formalism, which goes beyond the scope of this work. We thus limit ourselves to estimating the phenomenological impact of the KPCs inclusion, leaving the full scale determination of TMD distributions for a future study.

Since the current set of available TMD distributions contains only the unpolarized ones, our analysis is limited to the unpolarized structure functions. Furthermore, we neglect contributions from the Boer-Mulders and Collins functions, as they remain unknown. Based on our experience with Drell-Yan angular coefficients \cite{Piloneta:2024aac}, we estimate that the corrections from these functions are negligible for the present comparison. Among the unpolarized structure functions, the angle-integrated cross-section is the most relevant, since it serves as the normalization for polarized measurements (see, f.i., \cite{COMPASS:2016led, HERMES:2020ifk}). For this reason, we concentrate on it. 

The unpolarized SIDIS cross-section consists of two structure functions, $F_{UU,T}+\varepsilon F_{UU,L}$. The structure function $F_{UU,L}$ is power-suppressed and therefore neglected in all modern fits. It is naturally revealed in the TMD-with-KPCs formalism. To assess the significance of KPCs, we compare three scenarios: (i) the LP expression for $F_{UU,T}$, which is, so far, used to determine TMD distributions, (ii) the TMD-with-KPCs expression for $F_{UU,T}$, and (iii) the complete TMD-with-KPCs sum, $F_{UU,T}+\varepsilon F_{UU,L}$. In the figures \ref{fig:1}, \ref{fig:2}, \ref{fig:3}, we illustrate the comparison of (ii) (solid lines) and (iii) (dashed lines) against (i) for different kinematic configurations.

In fig.~\ref{fig:1} we present the KPCs magnitude as a function of $Q$. The comparison is intentionally performed at small $p_\perp$ to avoid contamination from $q_T/Q$ corrections. At $Q=2$GeV, the corrections are comparable in size to the LP contribution, both for $F_{UU,T}$ and $F_{UU,L}$ independently. The corrections then decrease rapidly: at $Q\sim 5$GeV, they amount to about 10-20\% (depending on other variables), and at $Q\sim10$GeV reduce to
$2-6$\%. Below $Q=2$GeV, power corrections become too large, and destroy the predictive power of factorization theorem.

\begin{figure}
\begin{center}
\includegraphics[width=0.98\textwidth]{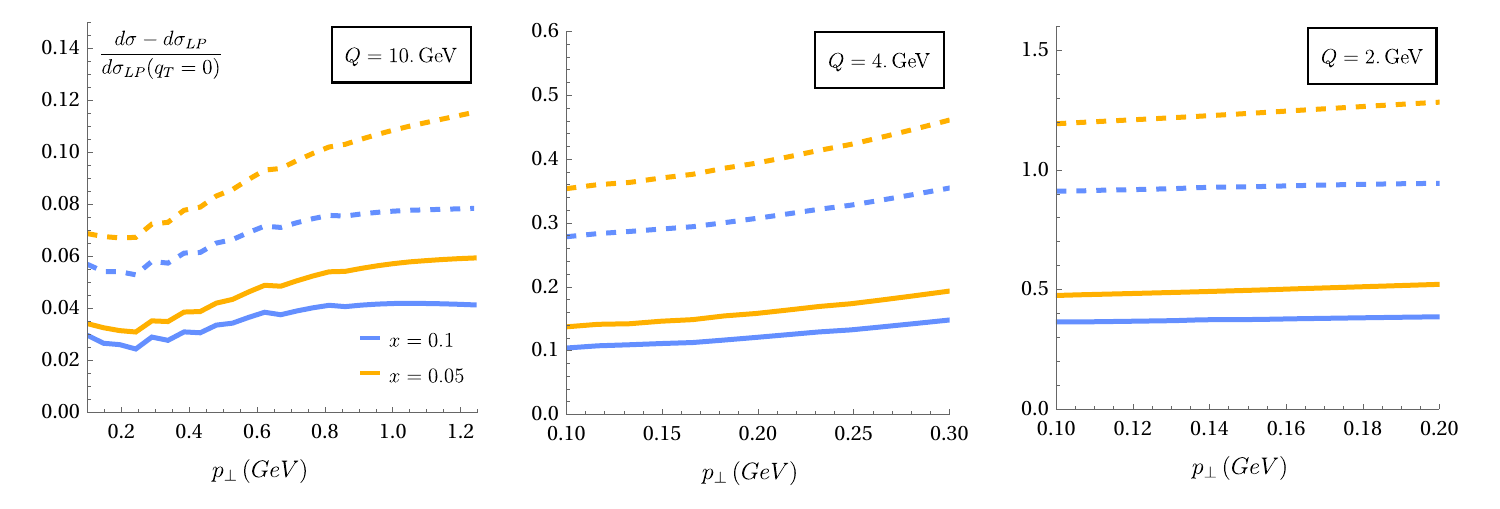}
\caption{\label{fig:2} Comparison of TMD-with-KPCs prediction to the pure LP TMD factorization expression for angle-integrated SIDIS cross-section as a function of $p_\perp$. Solid lines indicate the comparison of $F_{UU,T}$ only, while dashed lines include the contribution $+\varepsilon F_{UU,L}$ term. The comparison is done for $z=0.5$ and values of $(x,Q)$ are indicated in the figure. The value of $\varepsilon$ is defined by (\ref{def:varepsilon}) and is $\varepsilon\sim 0.85-1.$ for the present kinematics.}
\end{center}
\end{figure}

In fig.~\ref{fig:2} we show the KPCs magnitude as a function of $p_\perp$. The corrections for $F_{UU,T}$ and $F_{UU,L}$ exhibit a weak dependance on $p_\perp$, with a mild tendency to increase by about $2-5$\% for each. This effect is more pronounced for the sum $F_{UU,T}+\varepsilon F_{UU,L}$. Notably, the $F_{UU,L}$ contribution does not vanish at $p_\perp\to 0$, highlighting that KPCs remain significant within the traditional domain of the LP TMD factorization approach. From these plots, we found that the general size of power corrections relative to the LP expression can be estimated using the following pocket formulas
\begin{eqnarray}
\frac{F_{UU,T}}{F_{UU,T(LP)}}\sim 1+\frac{(2.3\text{GeV})^2}{Q^2}\frac{z^3}{x^{0.25}},
\qquad
\frac{F_{UU,L}}{F_{UU,T(LP)}}\sim \frac{(1.7\text{GeV})^2}{Q^2}\frac{z^{3/2}}{x^{0.2}}.
\end{eqnarray}
These expressions are not precise but provide an approximate order-of-magnitude estimation of the effects and can be used as a criterion to select experimental data. 

In these figures, small discontinuities in the curves are observed around $Q\sim 1$ and $Q\sim 4$GeV. They are related to the mass thresholds of the charm and top quarks, respectively. There are also instabilities in the curves with respect to $p_\perp$ at $Q=10$GeV (left panel of fig.~\ref{fig:2}), which arise from the computation of the LP term in \texttt{artemide} at $q_T\sim0$ and large-$Q$, where the Hankel transform degenerates and the algorithm becomes noisy. Nicely, the TMD-with-KPCs convolution is free from these numerical issues, as its integrand is not osculatory.

Finally, in fig.~\ref{fig:3} we present the computation of the SIDIS angle-integrated structure functions for the kinematics of current and future experiments. For comparison, we selected a high-energy bin from the JLab measurements \cite{JeffersonLabHallA:2016ctn}, a typical bin from the COMPASS measurements \cite{COMPASS:2017mvk}, and a low energy bin for the future EIC measurements \cite{AbdulKhalek:2021gbh}. In all these cases, the contribution of KPCs is significant, reaching 10-30\% at $x\sim 0.2$. The contribution from $F_{UU,L}$ is also substantial and comparable to the precision of experimental data.

\begin{figure}
\begin{center}
\includegraphics[width=0.98\textwidth]{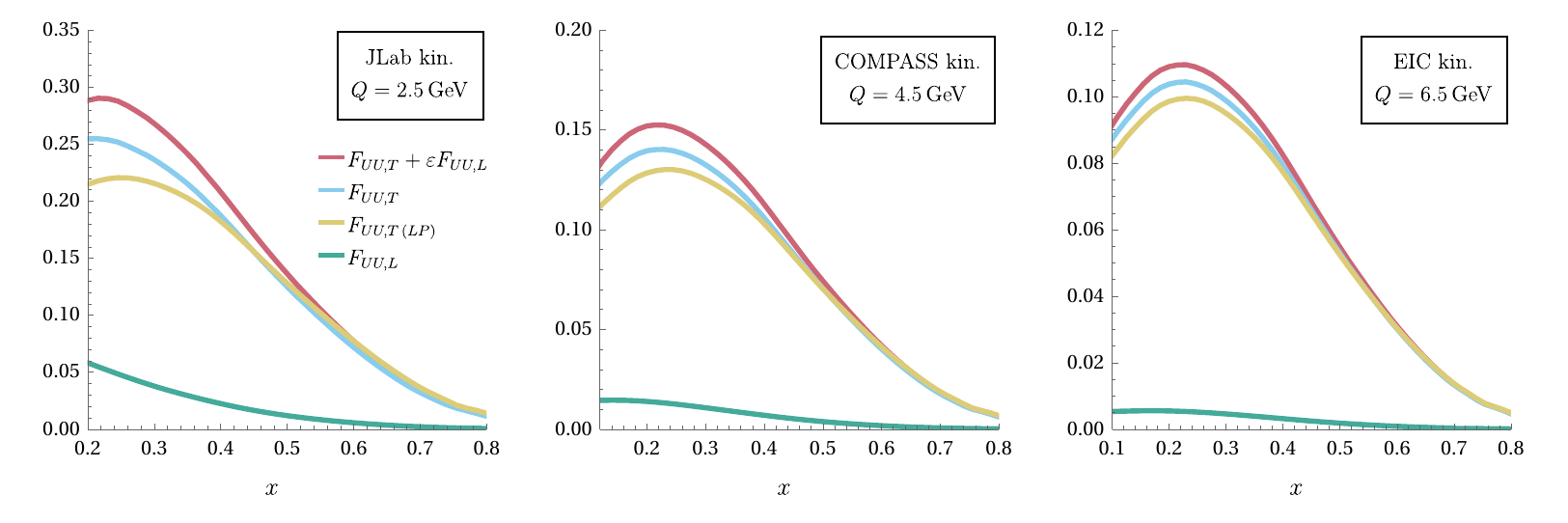}
\caption{\label{fig:3} Comparison of the TMD-with-KPCs predictions for the angle-integrated SIDIS structure functions $F_{UU,T}$ and $F_{UU,L}$ as a function of $x$, for the kinematics of JLab (left), COMPASS (center) and EIC (right), relative to the LP TMD factorization result for the structure function $F_{UU,T}$. The combination $F_{UU,T}+\varepsilon F_{UU,L}$ is also shown to better illustrate the contribution from longitudinal photons. The value of $\varepsilon$ is defined by (\ref{def:varepsilon}) and ranges approximately as $\varepsilon\sim 0.6-0.9$ for the JLab bin, as $\varepsilon\sim 0.5-0.9$ for the COMPASS bin, and as $\varepsilon\sim 0.8-1.$ for the EIC bin.}
\end{center}
\end{figure}

These observations lead us to conclude that the effect of power corrections is meaningful for the present SIDIS data. This effect is most pronounced in the normalization of the curve, which can change by a factor of 2 or 3 for lower-$Q$ bins, rather than in its shape, which varies only about a few percents. Consequently, the existing extractions of TMD distributions can not be considered reliable, and must be revisited taking into account the power corrections effects. It is well-known that, currently, there is disagreement among different groups regarding the interpretation of low energy SIDIS data (see, e.g., \cite{Moos:2025sal, Scimemi:2019cmh, Bacchetta:2022awv, Bacchetta:2024qre}). This disagreement stems from distinct readings of the LP formula at low energy, and it is possible that it could be mitigated by explicitly including KPCs.

\section{Conclusions}\label{sec:conclusions}

In this work, we have studied the structure functions that parametrize the semi-inclusive deep inelastic scattering (SIDIS) cross-section within the framework of the TMD-with-KPCs factorization theorem introduced in ref.~\cite{Vladimirov:2023aot}. This approach extends the standard transverse momentum dependent (TMD) factorization formalism by incorporating kinematic power corrections (KPCs), specifically those that follow the leading power (LP) term (see fig.~\ref{fig:power_pyramid}). As a result, it systematically includes all power-suppressed terms containing TMD distributions of twist-2 at small values of $q_T$. 

The inclusion of KPCs restores two fundamental properties of the hadron tensor that are violated in the LP approximation: charge conservation and frame invariance. Moreover, these corrections share both the perturbative content (i.e., the coefficient functions) and the non-perturbative content (i.e., the TMD distributions and the Collins-Soper kernel) of the LP factorization theorem, and can therefore be considered as an inherent part of it. This formalism is especially appealing because it has proven useful for computing observables that were previously theoretically inaccessible due to power-suppression; for example, the angular distributions of Drell-Yan (DY) leptons \cite{Piloneta:2024aac} or, in the case of SIDIS, the longitudinally polarized photons contributions.

The KPCs account for only a subset of the power corrections contributing to SIDIS at low-Q. Nevertheless, it is plausible that they constitute the dominant ones in the limit $p_\perp\ll Q$. In this regime, $q_T/Q$ corrections are, by definition, suppressed. The next possible contributions come from genuine twist-three distributions (denoted as $\Lambda/Q$ corrections in fig.~\ref{fig:power_pyramid}). However, existing studies consistently indicate that these corrections are much smaller than the twist-two ones. Therefore, it is reasonable to argue that the TMD-with-KPCs formalism captures the most relevant power corrections in this kinematic region.

We apply this methodology to compute the SIDIS hadron tensor in its general form, and, subsequently, for the structure functions. The main result of this paper consists of the expressions for the polarized SIDIS structure functions presented in sec.~\ref{subsec:StructFunctions}. They take the form of scattering cross-sections of massless on-shell partons, accompanied by rather cumbersome kinematic factors, arising mainly from the transformation between the Breit frame and the laboratory frame. The principal modification introduced by KPCs into the factorization formula is the restriction of the integration region over parton momenta to a finite domain.

With the help of the \texttt{artemide} code, we have conducted a series of impact studies to evaluate the phenomenological implications of including KPCs into unpolarized structure functions, and we have found that they significantly contribute at $Q<5-10$GeV, reaching up to 100\% already at $Q\sim 3$GeV. Furthermore, in this region, the contribution of longitudinally polarized photons becomes as large as the LP term. Consequently, incorporating power corrections is crucial for the description of all existing SIDIS data, most of which lies below $Q<5$GeV.

The numerical tests were performed using TMD distributions extracted within the LP factorization theorem \cite{Moos:2025sal}. Given the magnitude of these corrections, the reliability of these and other TMDFFs extractions is questionable. This finding can possibly shed light into the discrepancies among the different determinations of TMD distributions, as well as the tensions observed in the Collins-Soper kernel extractions. Consequently, the TMD distributions determinations should be revisited using an updated framework; an analysis we plan to perform in the future. 

\acknowledgments

We would like to thank Harut Avakian and Alexei Prokudin for useful comments and stimulating discussions. This work is funded by the \textit{Atracci\'on de Talento Investigador} program of the Comunidad de Madrid (Spain) No. 2020-T1/TIC-20204, and the grant ``Europa Excelencia'' No. EUR2023-143460 funded by MCIN/AEI/10.13039/501100011033/ by the Spanish Ministerio de Ciencias y Innovaci\'on. This project is also supported by the European Union Horizon research Marie Skłodowska-Curie Actions – Staff Exchanges, HORIZON-MSCA-2023-SE-01-101182937-HeI, DOI: 10.3030/101182937.

\appendix
\section{Useful expression for SIDIS kinematic}
\label{app:frames_dictionary}

In this appendix we collect the expressions useful for handling SIDIS kinematics in the TMD factorization framework, expressed in terms of the standard variables
\begin{eqnarray}
    Q^2 = -q^2, \quad x = \frac{Q^2}{2(Pq)}, \quad y = \frac{(Pq)}{(Pl)}, \quad z = \frac{(P p_h)}{(P q)}.
\end{eqnarray}
The parameters associated with the $1/Q$ corrections are defined as
\begin{eqnarray}
\gamma = \frac{2Mx}{Q}, \quad \gamma_h = \frac{m_h}{zQ}, \quad \gamma_\perp=\frac{\sqrt{\vec p_\perp^2}}{z Q}.
\end{eqnarray}
It should be pointed out that, in order to simplify the notation of the results outlined here, we also introduce the angle between $l$ and $q$ in the laboratory gamma-hadron frame \cite{Kotzinian:1994dv}
\begin{eqnarray}\label{app:theta-gamma}
\cos{\theta_\gamma} = \frac{1+y\frac{\gamma^2}{2}}{\sqrt{1+\gamma^2}} , \quad \sin{\theta_\gamma} = \frac{\gamma}{\sqrt{1+\gamma^2}}\sqrt{1-y-y^2\frac{\gamma^2}{4}}.
\end{eqnarray}
Some of the scalar products that appear most frequently throughout the calculations are
\begin{eqnarray}
(qp_h)&=&\frac{z Q^2}{\gamma^2}\(1-\sqrt{1+\gamma^2}\sqrt{1-\gamma^2 \gamma^2_h-\gamma^2 \gamma_\perp^2}\),
\\
(lp_h)&=&\frac{z Q^2}{y\gamma^2}\(1-\sqrt{1-\gamma^2 \gamma^2_h-\gamma^2 \gamma_\perp^2}\cos\theta_\gamma- \gamma \gamma_\perp \cos\phi_h \sin \theta_\gamma\),
\\
\epsilon^{\alpha\beta\gamma\delta}l_\alpha q_\beta P_\gamma p_{h\delta}&=&Q^4 \frac{z}{2 x y}\frac{\gamma_\perp}{\gamma}\sqrt{1+\gamma^2}\sin \phi_h\sin\theta_\gamma,
\end{eqnarray}
With their help, we found the following table of vector decompositions
\begin{eqnarray}\label{app:v1}
p_h^\mu&=&\frac{2xz}{\gamma^2}P^\mu +z \sqrt{\frac{1-\gamma^2\gamma_h^2-\gamma^2\gamma_\perp^2}{1+\gamma^2}}\(q^\mu-\frac{2x}{\gamma^2}P^\mu\)+p_\perp^\mu,
\\
q^\mu&=&\frac{2x}{\gamma^2}P^\mu +\frac{\sqrt{(1+\gamma^2)(1-\gamma^2\gamma_h^2-\gamma^2\gamma_\perp^2)}}{1-\gamma^2\gamma_h^2}\(\frac{p_h^\mu}{z}-\frac{2x}{\gamma^2}P^\mu\)+q_T^\mu,
\\
\tilde p_\perp^\mu &=& -z\sqrt{\frac{1-\gamma^2\gamma_h^2}{1+\gamma^2}}\tilde q_T^\mu,
\\
p_\perp^\mu&=& z\frac{\gamma^2 \gamma_\perp^2}{1-\gamma^2\gamma_h^2}\(\frac{p_h^\mu}{z}-\frac{2x}{\gamma^2}P^\mu\)
-
z\sqrt{\frac{1-\gamma^2\gamma_h^2-\gamma^2\gamma_\perp^2}{1+\gamma^2}}q_T^\mu,
\\\label{app:v5}
S^\mu &=&
S_\|\frac{\gamma}{Q}\(\frac{\sqrt{1-\gamma^2\gamma_h^2-\gamma^2\gamma_\perp^2}}{1-\gamma^2\gamma_h^2}\(\frac{2x}{\gamma^2}P^\mu -\frac{p_h^\mu}{z}\)-\frac{q_T^\mu}{\sqrt{1+\gamma^2}}\)
\\\nn &&-|s_\perp|\cos(\phi_h-\phi_S)\(\frac{\gamma_\perp}{Q}\frac{\gamma^2}{1-\gamma^2\gamma_h^2}\(\frac{2x}{\gamma^2}P^\mu-\frac{p_h^\mu}{z}\)+\sqrt{\frac{1-\gamma^2\gamma_h^2-\gamma^2\gamma_\perp^2}{1-\gamma^2\gamma_h^2}}\frac{q_T^\mu}{\sqrt{\vec q_T^2}}\)
\\\nn &&
-|s_\perp|\sin(\phi_h-\phi_S)\frac{\tilde q_T^\mu}{\sqrt{\vec q_T^2}}.
\end{eqnarray}
The meaning of the labels $\perp$ and $T$ that appear in the above expressions is explained in section \ref{subsec:SIDIS_kinematics}, where we introduce the notation of the two reference frame used in this work. 

Moreover, the relations between the normalizations of the spin components are
\begin{eqnarray}
|\vec s_T|=|\vec s_\perp|,\qquad \lambda=S_\|\sqrt{\frac{1-\gamma^2\gamma_h^2-\gamma^2\gamma_\perp^2}{1-\gamma^2\gamma_h^2}}- |\vec s_\perp|\cos(\phi_h-\phi_S) \frac{\gamma \gamma_\perp}{\sqrt{1-\gamma^2\gamma_h^2}}.
\end{eqnarray}

\subsection{Massless limit}
\label{app:massless}

In the massless limit, $M^2=m_h^2=0$, the vectors $P$ and $p_h$ reduces to
\begin{eqnarray}
P^\mu=\bar n^\mu P^+,\qquad p_h^\mu = n^\mu p_h^-,
\end{eqnarray}
with
\begin{eqnarray}
P^+p_h^-=\frac{z}{2x}Q^2.
\end{eqnarray}
The relation (\ref{rel:qT-to-pPerp}) between $p^2_\perp$ and $q_T^2$ simplifies
\begin{eqnarray}
\vec p_\perp^2=z^2\vec q_T^2,\qquad \gamma_\perp=\frac{\sqrt{\vec p_\perp^2}}{zQ}=\frac{\sqrt{\vec q_T^2}}{Q}.
\end{eqnarray}
The expressions for the vectors given in (\ref{app:v1}-\ref{app:v5}) turn into
\begin{eqnarray}
p_h^\mu&=&zq^\mu+z x P^\mu(1+\gamma_\perp^2)+p_\perp^\mu,
\\
q^\mu&=&\frac{p_h^\mu}{z}-xP^\mu(1-\gamma_\perp^2)+q_T^\mu,
\\
\tilde p_\perp^\mu &=& -z\tilde q_T^\mu ,
\\
p_\perp^\mu&=&-zq_T^\mu -2xz \gamma_\perp^2 P^\mu,
\\
S^\mu &=&
S_\|\frac{2x}{\gamma Q}P^\mu 
-|\vec s_\perp|\Big[\cos(\phi_h-\phi_S) \(\frac{2x\gamma_\perp}{Q}P^\mu+\frac{q_T^\mu}{\sqrt{\vec q_T^2}}\)
+\sin(\phi_h-\phi_S) \frac{\tilde q_T^\mu}{\sqrt{\vec q_T^2}}\Big],
\end{eqnarray}
and the relation between the spin componentssimply becomes $|\vec s_T|=|\vec s_\perp|$, $S_\|=\lambda$.

\section{Large-$Q$ limit of structure functions}
\label{app:largeQ}

In this appendix we collect the expressions for the structure functions in the large-$Q$ limit, which corresponds to the LP term of the conventional TMD factorization theorem. These expressions can be used for comparison with known results derived within the TMD factorization approach, and also as a minimal model for estimating structure functions whose LP contribution is absent. 

The structure functions presented below are the leading terms of the formal large-$Q$ expansion
\begin{eqnarray}
\lim_{\text{LP}}\mathcal{C}[A,f_1f_2]=\mathcal{C}_{\text{LP}}[\lim_{\text{LP}}A,f_1f_2],
\end{eqnarray}
where $\lim_{\text{LP}}A$ denotes the leading contribution in this limit. The integral convolution $\mathcal{C}_{\text{LP}}$ is 
\begin{eqnarray}\label{eq:convolution-LP}
\mathcal{C}_{\text{LP}}[A,f_1f_2] &=& F_0 ~x\sum_{f} e_f^2 ~C_{0,\text{DIS}}\(\frac{-2q^+q^-}{\mu^2}\) 
 \int d^2\vec k_{1T} d^2\vec k_{2T} \delta^{(2)}(\vec q_T+\vec k_{1T}-\vec k_{2T}) 
\\ \nn &&
\times (f_{1,f}(x_{\text{LP}},\vec k_1)f_{2,f}(z_{\text{LP}},\vec k_2)+\overline{f}_{1,f}(x_{\text{LP}},\vec k_1)\overline{f}_{2,f}(z_{\text{LP}},\vec k_2))A,
\end{eqnarray}
where
$$
x_{\text{LP}}=-\frac{q^+}{P^+},\qquad z_{\text{LP}}=\frac{p_h^-}{q^-}.
$$
It should be noted that we assume that TMD distributions decay faster than $1/\vec k^2$, which is approximately correct. This assumption allows us to evaluate separately the limits of the integration region and the integrand. The relative corrections to this expression are of order $Q^{-2}$, they include the modifications to the integration variables and boundaries, and can be expressed in terms of derivatives of TMD distributions as well as surface contributions. Consequently, the expression beyond the leading term for each structure function becomes rather involved. 

Below we provide the leading term expressions for each independent product of TMD distributions entering the structure functions. In some cases, the $Q$-counting of these terms does not coincide, and formally one would need to include higher order terms. However, as discussed above, this leads to cumbersome and practically useless expressions.
\begin{eqnarray}
F_{UU,T}&=&\mathcal{C}_{\text{LP}}[1,f_1D_1]+\frac{1}{2} F_{UU,L},
\\
F_{UU,L}&=&
\mathcal{C}_{\text{LP}}[\frac{4\vec k_1^2}{Q^2},f_1D_1]
+
\mathcal{C}_{\text{LP}}[-\frac{4\vec k_1^2}{Q^2}\frac{(\vec k_1\vec k_2)}{m_hM},h^\perp_1H^\perp_1]
,
\\
F_{UU}^{\cos\phi_h}&=&
\mathcal{C}_{\text{LP}}\[
\frac{2(\vec k_1\vec h_T)}{Q}
,f_1D_1\]
+
\mathcal{C}_{\text{LP}}\[
\frac{-2(\vec k_2\vec h_T)}{Q}\frac{\vec k_1^2}{m_hM}
,h^\perp_1H^\perp_1\]
,
\\
F_{UU}^{\cos2\phi_h}&=&-\frac{1}{2}F_{UU,L}+
\mathcal{C}_{\text{LP}}\[
\frac{4(\vec k_1 \vec h_T)^2}{Q^2}
,f_1D_1\]
+
\mathcal{C}_{\text{LP}}\Big[
\frac{(\vec k_1\vec k_2)-2(\vec k_1\vec h_T)(\vec k_2\vec h_T)}{m_hM}
,h^\perp_1H^\perp_1\Big],
\\
F_{UL}^{\sin\phi_h}&=&\mathcal{C}_{\text{LP}}\[-2\frac{(\vec k_2\vec h_T)}{Q}\frac{\vec k_1^2}{m_hM},h^\perp_{1L}H_{1}^\perp\],
\\
F_{UL}^{\sin2\phi_h}&=&\mathcal{C}_{\text{LP}}\Big[
\frac{(\vec k_1\vec k_2)-2(\vec k_1\vec h_T)(\vec k_2\vec h_T)}{m_hM},h^\perp_{1L}H_{1}^\perp\Bigg],
\\
F_{LL}&=&\mathcal{C}_{\text{LP}}\[1,g_{1L}D_{1}\],
\\
F_{LL}^{\cos\phi_h}&=&\mathcal{C}_{\text{LP}}\[2\frac{(\vec k_1 \vec h_T)}{Q},g_{1L}D_{1}\],
\\
F_{UT,T}^{\sin(\phi_h-\phi_S)}&=&\mathcal{C}_{\text{LP}}\[
\frac{(\vec k_1 \vec h_T)}{M}
,f^\perp_{1T}D_{1}\]+\frac{1}{2}F_{UT,L}^{\sin(\phi_h-\phi_S)},
\\
F_{UT,L}^{\sin(\phi_h-\phi_S)}&=&
\mathcal{C}_{\text{LP}}\[\frac{4(\vec k_1 \vec h_T)}{M}\frac{\vec k_1^2}{Q^2},f^\perp_{1T}D_{1}\]
+\mathcal{C}_{\text{LP}}\[\frac{2(\vec k_2\vec h_T)}{m_hM^2}\frac{(\vec k_1^2)^2}{Q^2},h^\perp_{1T}H^\perp_{1}\]
\\\nn &&
+\mathcal{C}_{\text{LP}}\[\frac{-4}{m_h}\frac{\vec k_2^2(\vec k_1\vec h_T)-(\vec k_1\vec k_2)\sqrt{\vec q_T^2}}{Q^2},h_{1}H^\perp_{1}\],
\\
F_{UT}^{\sin(\phi_h+\phi_S)}&=&
\mathcal{C}_{\text{LP}}\[
-\frac{(\vec k_1\vec h_T)}{M}\frac{\vec k_1^2}{Q^2},f^\perp_{1T}D_{1}\]
+
\mathcal{C}_{\text{LP}}\[\frac{(\vec k_2\vec h_T)}{m_h},h_{1}H^\perp_{1}\]
,
\\
F_{UT}^{\sin(3\phi_h-\phi_S)}&=&
\mathcal{C}_{\text{LP}}\[
\frac{4(\vec k_1\vec h_T)^2(\vec k_2\vec h_T)-4(\vec k_1\vec h_T)(\vec k_1\vec k_2)-\vec k_1^2(\vec k_1\vec h_T)}{M Q^2},f^\perp_{1T}D_{1}\]
\\\nn &&+
\mathcal{C}_{\text{LP}}\[\frac{4(\vec k_1\vec h_T)^2(\vec k_2\vec h_T)-2(\vec k_1\vec h_T)(\vec k_1\vec k_2)-\vec k_1^2(\vec k_2\vec h_T)}{2m_hM^2},h_{1T}^\perp H^\perp_{1}\]
\\
F_{UT}^{\sin(\phi_S)}&=&
\mathcal{C}_{\text{LP}}\[
\frac{-\vec k_1^2}{M Q},f^\perp_{1T}D_{1}\]
+
\mathcal{C}_{\text{LP}}\[\frac{2(\vec k_1\vec k_2)}{m_hQ},h_{1} H^\perp_{1}\]
\\
F_{UT}^{\sin(2\phi_h-\phi_S)}&=&
\mathcal{C}_{\text{LP}}\[
\frac{2(\vec k_1 \vec h_T)^2-\vec k_1^2}{M Q},f^\perp_{1T}D_{1}\]
\\\nn &&+
\mathcal{C}_{\text{LP}}\[\frac{\vec k_1^2}{m_hM^2}\frac{(\vec k_1\vec h_T)^2+(\vec k_1\vec h_T)(\vec k_2\vec h_T)-\vec k_1^2}{Q},h_{1T}^\perp H^\perp_{1}\]
\\
F_{LT}^{\cos(\phi_h-\phi_S)}&=&\mathcal{C}_{\text{LP}}\[
\frac{-(\vec k_1 \vec h_T)}{M},g_{1T}D_{1}\]
\\
F_{LT}^{\cos(\phi_S)}&=&\mathcal{C}_{\text{LP}}\[
\frac{-\vec k^2_1}{MQ},g_{1T}D_{1}\]
\\
F_{LT}^{\cos(2\phi_h-\phi_S)}&=&
\mathcal{C}_{\text{LP}}\[
\frac{\vec k_1^2-2(\vec k_1 \vec h_T)^2}{M Q},g_{1T}D_{1}\]
\end{eqnarray}
The structure functions that scale as $\sim Q^0$ can be compared with the results obtained within the LP TMD factorization theorem \cite{Bacchetta:2006tn}, and are in agreement with them. The terms $\sim Q^{-1}$ agree with the NLP expressions in refs.~\cite{Ebert:2021jhy, Rodini:2023plb}. On the other hand, some NNLP terms, such as $F_{UU}^{\cos\phi_h}$, can be found in ~\cite{Bacchetta:2008xw}, and also agree with our results. This provide an independent check of our computation.

\section{Details on the convolution integral and its implementation in \texttt{artemide}}\label{app:convolution_integral}

The convolution integral appearing in (\ref{hadronTensor}) is a central element of the TMD-with-KPCs factorization framework, as it guarantees frame-independence and a proper treatment of parton dynamics. From the numerical perspective, however, it has a rather involved structure due to the complexity of the integrand and the complicated shape of the integration region (illustrated in fig.~\ref{fig:integration_domain}). Therefore, rewriting it in a form suitable for numerical realization is a crucial step. In this appendix, we outline its main features and describe its implementation in the \texttt{artemide} code \cite{artemide}. 

The convolution (\ref{eq:convolution}) can be reassembled as follows
\begin{eqnarray}
\mathcal{C}[A,f_1f_2] &=& \frac{x}{2z}F_0\,Q^2 \sum_{f} e_f^2 ~C_{0,\text{DIS}}\(\frac{Q^2}{\mu^2}\)\mathcal{I}[A,f_1f_2],
\end{eqnarray}
where
\begin{eqnarray}\label{app:INT-def}
\mathcal{I}[A,f_1f_2] &=& 4P^+p_h^-
\\\nn &&
\int d^4k_1 d^4k_2 \delta^{(4)}(q+k_1-k_2) \delta(k_1^2)\delta(k_2^2)\int d\xi \frac{d\zeta}{\zeta} \delta\(k_1^+-\xi P^+\)\delta\(k_2^--\frac{p_h^-}{\zeta}\)
\\\nn &&
\times (f_{1,f}(\xi,\vec k_1)f_{2,f}(\zeta,\vec k_2)+\overline{f}_{1,f}(\xi,\vec k_1)\overline{f}_{2,f}(\zeta,\vec k_2))A.
\end{eqnarray}
Here, the quantity $A$ is a scalar function depending on the transverse momenta $\vec k_1$, $\vec k_2$, and $\vec q_T$, as well as on the collinear momentum fractions of the partons, $\xi$ and $\zeta$. The latter variables, as well as the scalar products of the transverse vectors, can be expressed in terms of the squared transverse momenta $\vec k_1^2$ and $\vec k_2^2$ by means of the $\delta$-functions. Specifically,
\begin{eqnarray}
    \xi = \frac{x_1}{2}\(1 - \frac{\vec k_1^2}{\tau^2} + \frac{\vec k_2^2}{\tau^2}+\frac{\sqrt{\lambda(\vec k_1^2, \vec k_2^2, -\tau^2)}}{\tau^2}\), 
    \, 
    \zeta = \frac{z}{2}\(1 - \frac{\vec k_1^2}{\vec k_2^2} - \frac{\tau^2}{\vec k_2^2}+\frac{\sqrt{\lambda(\vec k_1^2, \vec k_2^2, -\tau^2)}}{\vec k_2^2}\),
\end{eqnarray}
\begin{eqnarray}
(\vec k_1\vec k_2)=\frac{\vec k_1^2+\vec k_2^2-\vec q_T^2}{2},
\qquad
(\vec k_1\vec q_T)=\frac{\vec k_2^2-\vec k_1^2-\vec q_T^2}{2},
\qquad
(\vec k_2\vec q_T)=\frac{\vec k_2^2-\vec k_1^2+\vec q_T^2}{2},
\end{eqnarray}
where $\lambda$ denotes the triangle function, i.e.,
\begin{eqnarray}
    \lambda(a, b, c) = a^2 + b^2 + c^2 - 2ab - 2ac - 2bc,
\end{eqnarray}
and $\tau^2$ represents the hard-scale of the TMD factorization theorem
\begin{eqnarray}
\tau^2 = -2q^+q^- = Q^2 - \vec q_T^2.  
\end{eqnarray}
The variables $x_1$ and $z_1$ correspond to the LP expressions for the collinear momentum fractions,
\begin{eqnarray}
x_1& = &-\frac{q^+}{P^+},  
\qquad
z_1 = \frac{p_h^-}{q^-}.
\end{eqnarray}
In the massless limit, which is the case considered in this work, they simplify to
\begin{eqnarray}
M,m_h\to0~;\qquad  x_1 = x\frac{\tau^2}{Q^2}=x\(1-\frac{\vec q_T^2}{Q^2}\), \quad z_1 = z.
\end{eqnarray}
Moreover, we can exploit the $\delta$-functions to rewrite the integral (\ref{app:INT-def}) in a more convenient fashion
\begin{eqnarray}\label{app:INT2}
\mathcal{I}[A,f_1f_2] &=& \int_{R_{\xi\zeta}\cap R_T}d^2\vec k_{1T} d^2\vec k_{2T} \delta^{(2)}(\vec q_T+\vec k_{1T}-\vec k_{2T}) 
\\\nn &&
\times \frac{2\zeta}{\sqrt{\lambda(\vec k_1^2, \vec k_2^2, -\tau^2)}} (f_{1,f}(\xi,\vec k_1)f_{2,f}(\zeta,\vec k_2)+\overline{f}_{1,f}(\xi,\vec k_1)\overline{f}_{2,f}(\zeta,\vec k_2))A,
\end{eqnarray}
which facilitates comparison with the LP case by explicitly revealing the conventional $\delta$-function for transverse momenta. This also makes the differences with the LP convolution (\ref{eq:convolution-LP}) more transparent. The minor distinction lies in Jacobian factor $\zeta/\sqrt{\lambda}$ that incorporates power-suppressed contributions, while the more change concerns the integration region $R_{\xi\zeta}\cap R_T$. The region $R_{\xi\zeta}$ arises from the causality-imposed constraints $0 < \xi < 1$ and $0 < \zeta < 1$ applied to the TMD distributions. Explicitly, the region resulting from these conditions reads
\begin{eqnarray}
    R_{\xi\zeta}: \qquad \vec k_1^2 > 0 \,, \qquad 0 < \vec k_2^2 < \frac{(1-x_1)}{x_1}(x_1\vec k_1^2+\tau^2). 
\end{eqnarray}
The transverse $\delta$-function, meanwhile, constrains the integration domain as follows
\begin{eqnarray}
R_T : (|\vec q_T| - |\vec k_1|)^2 < \vec k_2^2 < (|\vec q_T| + |\vec k_1|)^2. 
\end{eqnarray}
Therefore, the integration domain is defined as the intersection $R_{\xi\zeta}\cap R_T$. Both regions are illustrated in fig.~\ref{fig:integration_domain}, highlighting another key difference between the LP and KPCs cases. 

As a consequence, the integrand acquires a more intricate structure, making the numerical evaluation of the convolution significantly more demanding. This also requires transforming the TMD distributions into momentum space. A fast and accurate algorithm for performing the corresponding Fourier-Bessel transformations is presented in the appendix B.2 of ref.~\cite{Piloneta:2024aac}.

To recover the LP case, one must take the limit $Q \to \infty$, where $\lambda(\vec k_1^2, \vec k_2^2, -\tau^2) \rightarrow Q^4$, and $\xi$ and $\zeta$ reduce to their LP counterparts, $x_1$ and $z$, respectively. The integrand then reduces to
\begin{eqnarray}
    \mathcal{I}_{\text{LP}}[\lim_{\text{LP}}A,f_1f_2]&=&\frac{2z}{Q^2}\int d^2\vec k_{1T} d^2\vec k_{2T} \delta^{(2)}(\vec q_T+\vec k_{1T}-\vec k_{2T})
    \\\nn &&
    \times (f_{1,f}(x_1,\vec k_1)f_{2,f}(z,\vec k_2)+\overline{f}_{1,f}(x_1,\vec k_1)\overline{f}_{2,f}(z,\vec k_2))\lim_{Q\to\infty}A.
\end{eqnarray}
Furthermore, in the LP limit, the integration boundaries $R_{\xi\zeta}$ become unbounded, and the first arguments of the distributions depend solely on $x_1$ and $z$. This allows to the convolution integral to be recast as a Hankel convolution in position space, in which the TMD factorization is used in the majority of modern applications.

In order to implement the integral (\ref{app:INT2}) within the \texttt{artemide} code and perform the numerical evaluation, it is more practical to reparametrize the variables in the integrand such that the integration region acquires a simple form. For this purpose, we introduce a new variable $\vec \Delta$ defined as
\begin{eqnarray}
\vec \Delta=-2\(\vec k_1+\frac{\vec q_T}{x_1}\).
\end{eqnarray}
The $\delta$-function fixes 
$$\vec k_1=-\frac{\vec \Delta}{2}-\frac{\vec q_T}{x_1},\qquad \vec k_2=-\frac{\vec \Delta}{2}-\frac{(1-x_1)\vec q_T}{x_1}.$$
The integration domain for $\vec \Delta$ is a circle of radius $\vec \Delta^2<4(1-x_1)Q^2/x_1^2$. Rescaling the norm of this new vector by $\sqrt{4(1-x_1)Q^2/x_1^2}$, the integral can be expressed over the unit disc
\begin{eqnarray}\label{app:INT3}
\mathcal{I}[A,f_1f_2] &=& \int_0^1 d\delta \int_0^{2\pi} d\theta \frac{2\xi}{\sqrt{\Lambda}}\frac{1-x_1}{x_1^2}\left(1+\frac{\vec q_T^2}{\tau^2}\right)\delta
\\\nn &&
\times (f_{1,f}(\xi,\vec k_1)f_{2,f}(\zeta,\vec k_2)+\overline{f}_{1,f}(\xi,\vec k_1)\overline{f}_{2,f}(\zeta,\vec k_2))A,
 \end{eqnarray}
Now, the relevant variables are defined as follows
\begin{align}
&\vec k_1^2=\frac{\tau^2}{4}(\Lambda-(1-S)^2), && \vec k_2^2=\frac{\tau^2}{4}(\Lambda-(1+S)^2),
\\\nn
& \xi=\frac{x_1}{2}(1-S+\sqrt{\Lambda}), && \zeta_h=\frac{2z}{1+S+\sqrt{\Lambda}},
\end{align} 
where 
\begin{eqnarray}
S &=&\frac{\vec k_1^2-\vec k_2^2}{2}= \frac{1}{\tau^2}\left(\frac{2-x_1}{x_1}\vec q_T^2 + \frac{2}{x_1}\delta\cos{\theta}\sqrt{\vec q_T^2(\tau^2+\vec q_T^2)(1-x_1)}\right), 
\\
\Lambda &=&
\frac{\lambda(\vec k_1^2,\vec k_2^2,-\tau^2)}{\tau^4}=\(
1 +\frac{\tau^2}{\vec q_T^2}S^2+4\frac{1-x_1}{x_1^2}\delta^2 \sin^2\theta \)\left(1+\frac{\vec q_T^2}{\tau^2}\right).
\end{eqnarray}
In this form the integral is sufficiently straightforward to compute. 

In the new variables, the expressions for the structure functions take rather lengthy forms. Essentially, they are polynomials in $S$ and $\Lambda$ with various coefficients. We do not present them here, as they carry no fundamental meaning, and can be directly extracted from the \texttt{artemide} code if needed.

When evaluating the integral (\ref{app:INT3}), it is important to note that the shape of the integrand exhibits a peak near $\delta\sim \sqrt{\vec q^2_T}/Q$ with a width $\sim Q/\sqrt{\vec q_T^2}$. This behavior results from the competing trends of the TMDPDF and TMDFF: while one grows at the border of the region, the other decreases, and vice-versa. Therefore, to address this, we found it convenient to further change the profile of the integral over $\delta$ such that the peaking region is zoomed, and thus the integration converges faster and more accurately.

\bibliography{bibFILE}

\end{document}